\documentclass[lettersize,journal]{IEEEtran}
\usepackage{amsmath,amsfonts}
\usepackage{algorithmic}
\usepackage{algorithm}
\usepackage{array}
\usepackage[caption=false,font=normalsize,labelfont=sf,textfont=sf]{subfig}
\usepackage{textcomp}
\usepackage{stfloats}
\usepackage{url}
\usepackage{verbatim}
\usepackage{graphicx}
\usepackage[nocompress]{cite}
\usepackage{amsmath,amssymb,amsfonts}
\usepackage{amsthm}
\usepackage{algorithmic}
\usepackage{algorithm}
\usepackage{graphicx}
\usepackage{textcomp}
\usepackage{xcolor}
\usepackage{graphicx}
\usepackage{bm,tikz,pgfplots,stmaryrd,wrapfig}
\usepackage{booktabs} 
\usepackage{multirow}
\DeclareMathOperator*{\argmax}{arg\,max}
\DeclareMathOperator*{\argmin}{arg\,min}
\usetikzlibrary{positioning,arrows,arrows.meta,shapes.geometric,calc,decorations.pathreplacing}
\tikzstyle{block} = [draw, rectangle, minimum height=1em, minimum width=2em,text centered]
\tikzstyle{lblock} = [draw, rectangle, minimum height=2em, minimum width=5cm,text centered]
\tikzstyle{summer} = [draw, circle, text centered]
\tikzstyle{attenuatorR} = [draw, regular polygon, regular polygon sides=3,shape border rotate=30, text centered]
\tikzstyle{antenna} = [draw, regular polygon, regular polygon sides=3,shape border rotate=180, text centered]

\def\BibTeX{{\rm B\kern-.05em{\sc i\kern-.025em b}\kern-.08em
    T\kern-.1667em\lower.7ex\hbox{E}\kern-.125emX}}

 \pgfplotsset{compat=1.18}
\begin{document}

\title{\fontsize{22}{26}\selectfont Sensing and Mitigation of Multi-Scatterer Self-Interference for Full-Duplex MIMO Communications}

\author{Anil~Kurt, \IEEEmembership{Graduate Student Member,~IEEE,} and Gokhan~M.~Guvensen, \IEEEmembership{Member,~IEEE}
\thanks{The authors are with the Department of Electrical and Electronics Engineering, Middle East Technical University, Ankara, Turkey (e-mail:
anilkurt@metu.edu.tr; guvensen@metu.edu.tr).}%
\thanks{Some initial concepts from this paper were motivated by our earlier conference paper \cite{Kurt_pimrc}, where only a single scatterer self-interference (SI) channel model is considered. On the other hand, this paper considers a multiple scatterer-based SI model and a more comprehensive framework, consisting of a selection policy of actions for SI mitigation.}
}

\markboth{}%
{}




\maketitle

\begin{abstract}
This paper proposes the joint use of digital self-interference cancellation (DSIC) and spatial suppression to mitigate far-field self-interference (SI) in full-duplex multiple-input multiple-output (MIMO) systems. Far-field SI, caused by echoes from environmental scatterers, is modeled based on the scatterers' angle and delay parameters, stored in a scatterer map. For each scatterer, the most suitable action regarding communication is selected from transmit beamforming, receive beamforming, DSIC, and no-action. This selection is based on simple metrics that show the expected uplink and downlink communication performance. In addition, emerging scatterers that deteriorate the communication are detected, and their delay and angles are acquired, providing an up-to-date scatterer map and presenting a \emph{sensing for communication} case. The proposed selection policy is compared with the individual implementations of DSIC and spatial suppression, highlighting the failure cases for each. It is shown that the proposed policy stays unaffected in these problematic cases and achieves SI-free performance.
\end{abstract}


\section{Introduction}
\IEEEPARstart{F}{ull-duplex} (FD) communication systems enable simultaneous transmission and reception using the same frequency and time resources \cite{Mohammadi_TenYears}. Besides potentially doubling the data rate, FD plays a significant role in many 6G applications, such as integrated sensing and communication (ISAC) \cite{Liu_FD_ISAC} and integrated access and backhaul (IAB) \cite{smida23}. However, FD transmission introduces several types of interference \cite{Sheemar_Multicell}. Among these, self-interference (SI), the leakage from transmitting antennas to receiving antennas within a transceiver, is a bottleneck for FD communications \cite{Sabharwal_Challenges}.  

SI can arise from direct leakage between the transmit (Tx) and receive (Rx) units, referred to as near-field SI (NFSI). Additionally, echoes from environmental scattering objects may cause far-field SI (FFSI) \cite{Kim_PHY_MAC,Sattari_Estimation}. While NFSI is strong and characterized by a time-invariant channel, FFSI is weaker due to greater path loss and is time-varying due to the movement of scatterers. For multiple-input multiple-output (MIMO) systems, NFSI spatial characteristics are challenging to model because of near-field effects \cite{Roberts_Codebooks, Roberts22}, whereas conventional radar signal models are applicable for FFSI \cite{Barneto_FDSensing, Liu_FD_ISAC}. Generally, SI mitigation techniques are classified as (a) passive methods and (b) active methods, where active methods further include (b.i) self-interference cancellation (SIC) and (b.ii) spatial suppression.

\subsection{Literature Review}
The existing literature primarily focuses on NFSI mitigation. Since NFSI is a strong interference ($\sim$104 dB above the signal of interest (SoI) \cite{Kim_PHY_MAC}), passive and active methods are applied in sequence. Regarding passive measures, single-antenna systems exploit suppression via circulator circuitry \cite{smida23}, while multi-panel systems employ techniques such as antenna separation, directional isolation, cross-polarization, absorptive shielding \cite{Everett14, Chen_propagation}, and even movable antennas \cite{Ding_Movable}. 

Following passive measures, SIC is the primary active method, involving the subtraction of an SI replica from the received signal \cite{Askar_Handling}. The high strength of SI makes it impossible to digitize alongside the weaker SoI due to the limited dynamic range of analog-to-digital converters (ADCs) \cite{Sabharwal_Challenges}. Consequently, analog SIC is implemented using additional radio frequency (RF) circuits \cite{Le_BeamBasedASIC, Kwak_ASIC, Kim_Duplexer}. However, analog SIC has limitations as its input cannot be digitally observed, necessitating adaptive methods (e.g., LMS) that rely on output observation \cite{Soriano_Adaptive, Le_BeamBasedASIC}. Moreover, the models used must accommodate multiple taps (due to the dispersion effect of fractional delays) \cite{Kiayani_NLRF, Kwak_ASIC, Le_BeamBasedASIC, Liu_ASICwImperfect} and account for nonlinearities (e.g., power amplifier effects, I/Q imbalance) using models such as memory polynomials (MP) \cite{Liu_dig_assisted, Kiayani_NLRF, kurt23}. Due to these difficulties, analog SIC remains suboptimal, often leaving residual interference, though it facilitates digitization.

Digital SIC (DSIC) can use more inclusive nonlinear models, such as generalized memory polynomials (GMP) \cite{kurt23} and B-spline interpolation \cite{Campo_Spline}. Despite its nonlinear nature, NFSI is accurately modeled due to its time-invariant characteristics, enabling error minimization through time integration to mitigate noise effects. Additionally, machine learning techniques can leverage a large number of iterations during learning \cite{Elsayed_ML}. The primary design challenge lies in managing the physical or computational complexity of SI regeneration. As highlighted, nonlinear modeling techniques in both analog and digital domains dominate the literature on SI mitigation.

Multipanel MIMO systems are resilient to NFSI as the panels remain outside each other's beams. Although analog SIC has been studied \cite{Le_BeamBasedASIC}, millimeter-wave measurements show a 40 dB interference-to-noise ratio (INR) even without active mitigation efforts \cite{Roberts22}, suggesting that analog SIC can be omitted. In the digital stage, spatial suppression is generally favored over DSIC, where beamformer weights are optimized based on a model or estimate of the spatial NFSI channel \cite{Barneto_FDSensing, Everett_SoftNull, kim23, Balti_HBFDesign}.

On the other hand, the literature on the FFSI is weaker. For single-antenna systems, a single SI channel is considered where larger delays represent time-varying FFSI components with Rayleigh or Rician fading \cite{Kwak_ASIC, Liu_ASICwImperfect, kurt23}. However, this results in a both nonlinear and time-varying SI channel. For MIMO systems, NFSI and FFSI channels are modeled separately due to spatial near-field effects on NFSI \cite{koc21,Balti_HBFDesign}. While \cite{koc21} considers angles of scatterers for FFSI, \cite{Balti_HBFDesign} introduces a time-varying model but scatterers are not discussed. Generally, the delay difference due to the distinct propagation distances is overlooked. Although more rigorous models are seen in monostatic ISAC studies \cite{Liu_FD_ISAC}, where FFSI is actually the intended signal, effects on communication are generally not the primary focus. In terms of mitigation, FFSI is generally handled with spatial suppression via beamformer optimization \cite{kim23,koc21,Balti_HBFDesign,Kurt_pimrc}. Comparing spatial suppression and DSIC for FFSI mitigation, \cite{kim23} discusses the effects of the number of antennas and spatial correlation, while \cite{Kurt_pimrc} discusses the effects of user/scatterer positioning.

\subsection{Motivation and Contributions}
Being weaker, FFSI is generally considered an appendix to NFSI. Studies on their joint mitigation show successful results and FFSI often appears ineffective as the focus is placed on nonlinearity rather than time variation. However, time variation limits the duration for both training and usage of models for this joint approach as shown in \cite{kurt23}, also restricting the period over which the reported performances remain valid. In addition, NFSI needs nonlinear modeling due to its strength but FFSI is similar in strength to SoI, for which linear models suffice. Therefore, treating FFSI and NFSI as two additive terms is more suitable due to differences in delay regions, spatial near-field effects, and nonlinear modeling needs. We can assume a nonlinear time-invariant channel for NFSI and a linear time-varying channel for FFSI, simplifying the mitigation stages for each. Separate mitigation removes the concerns about the time variation for well-studied NFSI mitigation methods and allows a more rigorous discussion on special features of FFSI mitigation, which is an often adopted idea in monostatic ISAC studies \cite{Keskin_Monostatic_PN, Liu_ISAC_toward}.

FFSI mitigation problem becomes more interesting for MIMO systems with the availability of spatial suppression besides SIC, bringing the geometry of transceivers and scatterers into the discussion. Although FFSI models are weak in related literature due to the aforementioned focus on NFSI, angle-based correlated spatial models are available for millimeter wave channels \cite{Sattari_Estimation, koc21}, known for their angle-delay sparsity \cite{Chen_Sparsity, Rangan_Sparsity, Sloane_Sparsity}. Together with the SI delay needed by SIC, the problem connects with the idea of acquisition and utilization of a scatterer map with related parameters.

Motivated by the comparison of spatial suppression and DSIC for single-scatterer FFSI in our previous work \cite{Kurt_pimrc}, this paper proposes the joint utilization of both methods for the case of multiple scatterers. For this purpose, NFSI, weakened by the multipanel design, is assumed to be mitigated by existing well-studied methods, while FFSI mitigation is designed as a system operating only when needed and leveraging suitable parameters. The contributions of this work are as follows:
\begin{itemize}
    \item The advantages and disadvantages of \emph{DSIC} and \emph{spatial suppression} for FFSI are investigated and compared. 
    \item A novel technique is offered for using DSIC and spatial suppression \emph{together} for the \emph{multi-scatterer} case, avoiding the disadvantages for each. For this, \emph{simple metrics} are derived for the expected communication performance in the case of each SI mitigation action, namely Tx beamforming, Rx beamforming, DSIC, and no-action. Based on these metrics, the optimal action for \emph{each scatterer} is selected to maximize communication performance.
    \item The requirements of these methods, namely the \emph{SI delay} for DSIC and the \emph{scatterer angle} for spatial suppression, are also considered and estimators are offered. In fact, the delay estimator is also a detector for \emph{a newly emerging scatterer}, which is also a considered case.
    \item Since the delays and angles of scatterers are estimated, stored in the \emph{scatterer map}, and utilized to enhance communication, this work serves as an example of \emph{sensing for communication}.
\end{itemize} 

\emph{Notation:} For an arbitrary matrix $\bm{X}$, $(\bm{X})_{r,c}$ gives the entry in $r$\textsuperscript{th} row and $c$\textsuperscript{th} column, $(\bm{X})_{r,:}$ is the $r$\textsuperscript{th} row, $(\bm{X})_{:,c}$ is the $c$\textsuperscript{th} column. In addition, $\bm{X}^{T}$ and $\bm{X}^{H}$ are the transpose and the conjugate transpose of the matrix $\bm{X}$.

\section{System Model}
\begin{figure}[bt]
\centering
\resizebox{0.40\textwidth}{!}{%
	\includegraphics{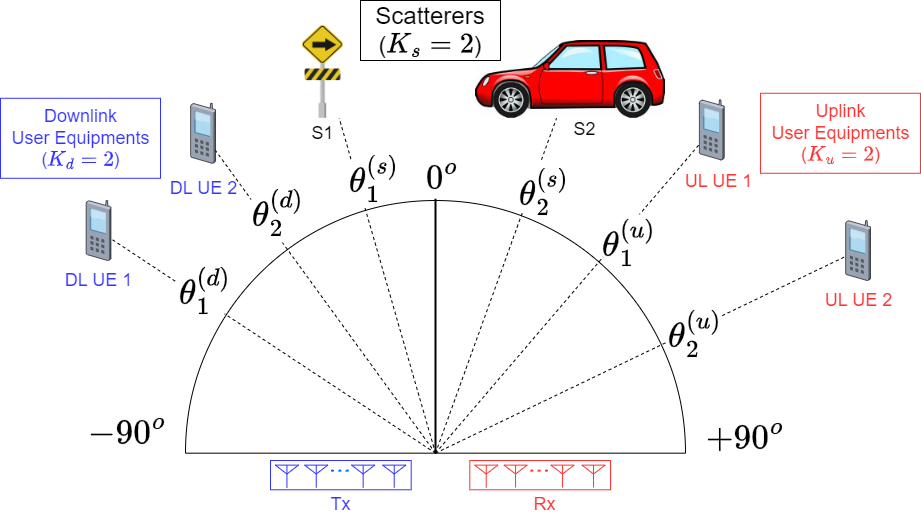}
	}
	\caption{Angular convention and scenario illustration for exemplary numbers $K_u$, $K_d$ and $K_s$.}
	\label{fig_angulardomain}
\end{figure}

\begin{figure*}[bt]
\centering
\resizebox{0.90\textwidth}{!}{%
\begin{tikzpicture}[>=triangle 45]

\def\sc{1.8}  
\def\ysc{1*\sc} 

\def\xdel{1.4*\sc}
\def\xa{0}
\def\xb{\xa + 0.3*\xdel}
\def\xc{\xb + \xdel}
\def\xd{\xc + \xdel}
\def\xe{\xd + \xdel}
\def\xf{\xe + \xdel}
\def\xg{\xf + \xdel}
\def\xh{\xg + 0.7*\xdel}
\def\xi{\xh + 1.2*\xdel}
\def\xj{\xi + 0.3*\xdel}
\def\xk{\xj + \xdel}
\def\xl{\xk + 0.7*\xdel}

\def\ydel{1*\ysc}
\def\ya{0}
\def\yb{\ya - \ydel}
\def\yc{\yb - 0.95*\ydel}
\def\yd{\yc - 0.95*\ydel}
\def\ye{\yd - 0.5*\ydel}

\def\eps{0.1*\sc}

\node (dd) at (\xb,\yb) {$\bm{D}^{(d)}$};
\node (dft) [block, text width=5em, minimum height=3em] at (\xc,\yb) {Multi-user\\Precoder};
\node (dzf) [block, minimum height=3em] at (\xe,\yb) {IFFT};
\node (dbf) [block, text width=3em] at (\xg,\yb) {Tx BF\\$\bm{W}$};
\node (dch) [block, minimum height=3em, text width=4em] at (\xj,\yb) {DL Ch.\\$\bm{H}^{(d)}$};
\node (dft2) [block, minimum height=3em] at (\xk,\yb) {FFT};
\node (dd_) at (\xl,\yb) {$\hat{\bm{D}}^{(d)}$};

\node (nfch) [block, minimum height=3em, text width=4em] at (\xh,\yc) {NFSI Ch.};
\node (sch) [block, minimum height=3em, text width=10em] at (\xi,\yc) {FFSI Channels\\$\bm{H}_k^{(s)}$ for $k \in \mathcal{S}$};

\node (ud_) at (\xb,\yd) {$\hat{\bm{D}}^{(u)}$};
\node (uzf) [block, minimum height=3em, text width=5em] at (\xc,\yd) {Multi-user\\Equalizer};
\node (uff) [block, minimum height=3em, text width=5em] at (\xd,\yd) {FFSI\\Mitigation};
\node (uft) [block, minimum height=3em] at (\xe,\yd) {FFT};
\node (unf) [block, minimum height=3em, text width=5em] at (\xf,\yd) {NFSI\\Mitigation};
\node (ubf) [block, minimum height=3em, text width=3em] at (\xg,\yd) {Rx BF\\$\bm{C}$};
\node (usum) [summer] at (\xh,\yd) {\tiny $\sum$};
\node (uch) [block, text width=4em, minimum height=3em] at (\xj,\yd) {UL Ch.\\$\bm{H}^{(u)}$};
\node (uft2) [block, minimum height=3em] at (\xk,\yd) {IFFT};
\node (ud) at (\xl,\yd) {$\bm{D}^{(u)}$};
\node (ztext) [anchor=south] at ($(uff.east)!0.5!(uft.west)$) {$\bm{Z}$};
\node (fdtext) [anchor=south] at ($(dft.east)!0.5!(dzf.west)$) {$\bm{F}^{(d)}$};
\coordinate (bottom) at (\xc,\ye) {};
\coordinate (bottom1) at (bottom-|ud_.west) {};
\coordinate (bottom2) at ($(bottom-|ubf.east)!0.5!(bottom-|nfch.west)$) {};
\coordinate (bottom3) at ($(bottom-|sch.east)!0.5!(bottom-|uft2.west)$) {};
\coordinate (bottom4) at (bottom-|ud.east) {};
\draw (bottom1) -- (bottom4);
\draw (bottom1) -- ($(bottom1)+(0,\eps)$);
\draw (bottom2) -- ($(bottom2)+(0,\eps)$);
\draw (bottom3) -- ($(bottom3)+(0,\eps)$);
\draw (bottom4) -- ($(bottom4)+(0,\eps)$);
\node () [fill=white] at ($(bottom1)!0.5!(bottom2)$) {Base Station};
\node () [fill=white] at ($(bottom2)!0.5!(bottom3)$) {Channel};
\node () [fill=white] at ($(bottom3)!0.5!(bottom4)$) {Users};

\coordinate (left) at (\xa,\yc) {};
\coordinate (left1) at (left|-dft.north) {};
\coordinate (left2) at (left|-dft.south) {};
\coordinate (left3) at (left|-uzf.north) {};
\coordinate (left4) at (left|-uzf.south) {};
\draw ($(left1)+(\eps,0)$) -- (left1) -- (left2) -- ($(left2)+(\eps,0)$);
\draw ($(left3)+(\eps,0)$) -- (left3) -- (left4) -- ($(left4)+(\eps,0)$);
\node () [fill=white, rotate=90] at ($(left1)!0.5!(left2)$) {DL};
\node () [fill=white, rotate=90] at ($(left3)!0.5!(left4)$) {UL};

\draw[->] (dd) -- (dft);
\draw[->] (dft) -- (dzf);
\draw[->] (dzf) -- (dbf);
\draw[->] (dbf) -- (dch);
\draw[->] (dch) -- (dft2);
\draw[->] (dft2) -- (dd_);
\draw[<-] (ud_) -- (uzf);
\draw[<-] (uzf) -- (uff);
\draw[<-] (uff) -- (uft);
\draw[<-] (uft) -- (unf);
\draw[<-] (unf) -- (ubf);
\draw[<-] (usum) -- (uch);
\draw[<-] (ubf) -- (usum);
\draw[<-] (uch) -- (uft2);
\draw[<-] (uft2) -- (ud);
\draw[->] (dbf-|usum) -- ($(sch.north west)!0.35!(sch.north east)$);
\draw[<-] (usum) -- ($(sch.south west)!0.35!(sch.south east)$);
\draw[->] (dbf-|usum) -- ($(sch.north west)!0.85!(sch.north east)$);
\draw[<-] (usum) -- ($(sch.south west)!0.85!(sch.south east)$);
\node () [anchor=south] at (sch.north) {$\dots$};
\node () [anchor=north] at (sch.south) {$\dots$};
\draw[->] (dbf) -| (nfch);
\draw[<-] (usum) -- (nfch);
\end{tikzpicture}
}
\caption{General system diagram to be reduced to the considered equivalent diagram in Fig. \ref{fig_sys_diag_eq}. (Ch.: Channel).}
\label{fig_sys_diag}
\end{figure*}
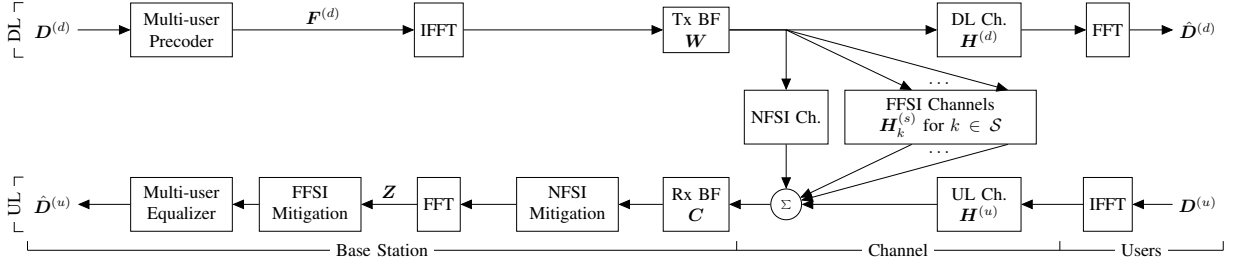

The considered FD system features a multipanel MIMO base station (BS), orthogonal frequency-domain multiplexing (OFDM), and millimeter-wave communication. $K_u$ uplink (UL) users, $K_d$ downlink (DL) users, and $K_s$ scatterers are considered as shown in Fig. \ref{fig_angulardomain}, forming the sets $\mathcal{U}_u \triangleq \{1,\dots,K_u\}$, $\mathcal{U}_d \triangleq \{1,\dots,K_d\}$, and $\mathcal{S} \triangleq \{1,\dots,K_s\}$, respectively. Also, a UL/DL user pair in the same position describes an FD user equipment. While the users have single-antenna systems, the BS consists of colocated Tx and Rx units with uniform linear arrays (ULAs) with $N_a$ antenna elements. The BS employs a hybrid beamformer (HBF) structure, where RF chains (RFCs) are fewer than antennas. Variables related to UL/DL users and scatterers are indicated by $u$, $d$, and $s$, respectively. For instance, the angles are denoted by $\theta_k^{(u)}$, $\theta_k^{(d)}$, and $\theta_k^{(s)}$ as in Fig. \ref{fig_angulardomain}, where $k$ is the member's index in the related group. Besides the inclusive system diagram in Fig. \ref{fig_sys_diag}, an equivalent diagram is given in Fig. \ref{fig_sys_diag_eq} that better reflects the considered system model.

\subsection{Transmitted Signals}
Symbols are denoted by $d_{k,m}^{(u)}$ for UL user $k \in \mathcal{U}_u$, and $d_{k,m}^{(d)}$ for DL user $k \in \mathcal{U}_d$, for subcarrier $m \in \{0,\dots,N_c-1\}$. UL users implement the inverse fast Fourier transform (IFFT) to obtain the time domain signal $s_{k,n}^{(u)}$ for time $n$ as\footnote{Cyclic prefix and oversampling are not shown for notational simplicity.}
\begin{equation}
    s_{k,n}^{(u)} = \frac{1}{\sqrt{N_c}} \sum_{m=0}^{N_c-1} d_{k,m}^{(u)} e^{j \frac{2\pi}{N_c}mn}, \qquad n=0,\dots,N_c-1,
\end{equation}
to be transmitted by the $k$\textsuperscript{th} UL user without further processing. 

For the DL communication, the BS first implements multi-user precoding to obtain $f_{k,m}^{(d)}$ from $d_{k,m}^{(d)}$. The precoding is described as $\bm{F}^{(d)} = \bm{T} \bm{D}^{(d)}$, where $\left(\bm{F}^{(d)}\right)_{km}=f_{k,m}^{(d)}$, $\left(\bm{D}^{(d)}\right)_{km}=d_{k,m}^{(d)}$, and $\bm{T}$ is $K_d \times K_d$ invertible multi-user precoding matrix, which will be detailed in Section \ref{sec_ch_eq}.
Then, the time domain signal $s_{k,n}^{(d)}$ is obtained via IFFT as
\begin{equation}
    s_{k,n}^{(d)} = \frac{1}{\sqrt{N_c}} \sum_{m=0}^{N_c-1} f_{k,m}^{(d)} e^{j \frac{2\pi}{N_c}mn}, \qquad n=0,\dots,N_c-1.
\end{equation}
Finally, BS transmits the beamformed signal $\bm{x}_n$, written as
\begin{equation}
    \bm{x}_n = \bm{W} \bm{s}_{n}^{(d)},
\end{equation}
 where $\bm{W} \triangleq [\bm{w}_1 \, \cdots \, \bm{w}_{K_d}]$ is $N_a \times K_d$ statistical Tx beamforming matrix and $\bm{s}_{n}^{(d)} \triangleq [s_{1,n}^{(d)} \, \cdots \, s_{K_d,n}^{(d)}]^{T}$. While $\bm{W}$ depends on the angles of the DL users, $\bm{F}$ depends on the instantaneous channels with much shorter coherence intervals. 

\subsection{Channels}
Channels have temporal and spatial characteristics. To describe the spatial aspects, the steering vector $\bm{a}(\theta)$ for a half-wavelength ULA with $N_a$ elements will be used as 
\begin{equation}
    \bm{a}(\theta) \triangleq \frac{1}{\sqrt{N_a}}\left[ 1 \quad e^{j \pi \sin{\theta}} \, \dots \, e^{j (N_a-1)\pi \sin{\theta}} \right]^{T},
\end{equation}
where $\theta$ is the angle as shown in Fig. \ref{fig_angulardomain}. When the angle is for scatterers or users, we will use
\begin{equation}
    \bm{q}_k^{(x)} \triangleq \bm{a}(\theta_k^{(x)}), \qquad x \in \{u,d,s\}.
\end{equation}

The spatial channels for the $k$\textsuperscript{th} DL user and the $k$\textsuperscript{th} UL user are denoted by $\bm{h}_k^{(d)}$ and $\bm{h}_k^{(u)}$, respectively, written as\footnote{Temporal spread (multipath) and angular spread effects are not considered due to the well-known angle-delay sparsity in the millimeter-wave channels \cite{Chen_Sparsity, Rangan_Sparsity, Sloane_Sparsity}. In addition, possible delays between all users are considered to be compensated as shown in \eqref{eq_Z} to focus on the effects of SI.}
\begin{equation}
    \bm{h}_k^{(x)} = \alpha_k^{(x)} \bm{q}_k^{(x)}, \qquad x \in \{u,d\},
\end{equation}
where $\alpha_k^{(x)} \sim \mathcal{CN}(0,P_k^{(x)})$. Reduced-dimensional effective channels $\bm{g}_k^{(d)}$ and $\bm{g}_k^{(u)}$ express these full-dimensional channels together with beamformers as
\begin{align}
    \bm{g}_k^{(u)} \triangleq \bm{C}^H \bm{h}_k^{(u)}, && \bm{g}_k^{(d)} \triangleq \bm{W}^H \bm{h}_k^{(d)}.
\end{align}
where $\bm{C} \triangleq [\bm{c}_1 \, \cdots \, \bm{c}_{K_u}]$ is the $N_a \times K_u$ Rx beamformer matrix. While $\bm{h}_k^{(x)}$ has length $N_a$,  $\bm{g}_k^{(x)}$ has length $K_x<N_a$ for $x \in \{u,d\}$.

There are also NFSI and FFSI channels, as shown in Fig. \ref{fig_sys_diag}. While NFSI mitigation is out of scope, the FFSI channels describe the reflection of the Tx signal $\bm{x}_n$ from the scatterers in the environment. The echo from the $k$\textsuperscript{th} scatterer is $\bm{H}_k^{(s)} \bm{x}_{n-l_k^{(s)}}^{(d)}$ with delay $l_k^{(s)}$ relative to UL signals. The $N_a \times N_a$ FFSI channel matrix $\bm{H}_k^{(s)}$ is defined as 
\begin{equation}
    \bm{H}_k^{(s)} \triangleq \alpha_k^{(s)} \bm{q}_k^{(s)} (\bm{q}_k^{(s)})^{H},
\end{equation}
with $\alpha_k^{(s)} \hspace{-1pt} \sim \mathcal{CN}(0,P_k^{(s)})$, depending on the scatterer angle $\theta_k^{(s)}$ through $\bm{q}_k^{(s)}$. After Rx beamformer, the FFSI is expressed as $\bm{G}_k^{(s)} \bm{s}_{n-l_k^{(s)}}^{(d)}$ with the effective FFSI channel $\bm{G}_k^{(s)}$, written as
\begin{align} \label{eq_Gks}
    \bm{G}_k^{(s)} = \alpha_k^{(s)} \bm{C}^{H} \bm{q}_k^{(s)} (\bm{q}_k^{(s)})^{H} \bm{W},
\end{align}
which is a system from $K_d$ DL streams to $K_u$ Rx chains.

\subsection{Received Signals}
The total signal that BS receives in time $n$ is
\begin{equation}
    \bm{r}_n =  \sum_{k=1}^{K_u} \bm{h}_k^{(u)} s_{k,n}^{(u)} 
             + \sum_{k=1}^{K_s} \bm{H}_k^{(s)} \bm{W} \bm{s}_{n-l_k^{(s)}}^{(d)} 
             + \bm{\eta}_n,
\end{equation}
where $\bm{\eta}_n \sim \mathcal{CN}(\bm{0},N_0 \bm{I}_{N_a})$ is the thermal noise. After Rx beamforming, $K_u \times 1$ vector $\bm{y}_n \triangleq \bm{C}^{H} \bm{r}_n$ is obtained, whose each entry is designated to a UL user. It is written as
\begin{align}
    \bm{y}_n = \sum_{k=1}^{K_u} \bm{g}_k^{(u)} s_{k,n}^{(u)} 
    + \sum_{k=1}^{K_s} \bm{G}_k^{(s)} \bm{s}_{n-l_k^{(s)}}^{(d)} 
    + \bm{C}^{H} \bm{\eta}_n.
\end{align}
After collecting $\bm{y}_n$ for $n=0,\dots,N_c-1$, information on subcarriers is extracted via fast Fourier transform (FFT). It is defined as $\bm{z}_m \triangleq \text{FFT} \{ \bm{y}_n \}_m = \frac{1}{\sqrt{N_c}} \sum_{n=0}^{N_c-1} \bm{y}_n e^{-j \frac{2\pi}{N_c} m n}$ for the subcarrier $m$, and expressed as
\begin{align}
    \bm{z}_m & = \sum_{k=1}^{K_u} \bm{g}_k^{(u)} d_{k,m}^{(u)} + \sum_{k=1}^{K_s} \bm{G}_k^{(s)} \bm{f}_{:,m}^{(d)} e^{-j \frac{2\pi}{N_c} m l_k^{(s)}} + \bm{\eta}_m^{(z)},
\end{align}
where $\bm{\eta}_m^{(z)} \sim \mathcal{CN}(\bm{0},N_0 \bm{C}^{H} \bm{C})$ is the noise term. Information from all the subcarriers is collected as columns of the $K_u \times N_c$ matrix $\bm{Z} \triangleq \left[ \bm{z}_1 \cdots \bm{z}_{N_c} \right]$, which is expressed as
\begin{equation} \label{eq_Z}
    \bm{Z} = \underbrace{\bm{G}^{(u)} \bm{D}^{(u)}}_{\text{SoI}} + \underbrace{\sum_{k=1}^{K_s} \bm{G}_k^{(s)} \bm{F}^{(d)} \bm{E}_{l_k^{(s)}}}_{\text{SI}} + \underbrace{\bm{N}}_{\text{Noise}},
\end{equation}
where 
$\bm{G}^{(u)} \triangleq \left[ \bm{g}_1^{(u)} \cdots \bm{g}_{K_u}^{(u)} \right]$, 
$( \bm{D}^{(u)} )_{k,m} \triangleq d_{k,m}^{(u)}$, 
$\bm{E}_l \triangleq \text{diag} \{e^{-j \frac{2\pi}{N_c} m l}\}_m$, and $\bm{N} \triangleq \left[ \bm{\eta}_1^{(z)} \cdots \bm{\eta}_{N_c}^{(z)} \right]$. 
As seen, the observation consists of SoI, SI, and noise terms.

For DL, the $k$\textsuperscript{th} user receives $(\bm{h}_k^{(d)})^{H} \bm{x}_n = (\bm{g}_k^{(d)})^{H} \bm{s}_{n}^{(d)}$. Each DL user collects signals from $n=0$ to $n=N_c-1$ and applies FFT to reach information in $N_c$ subcarriers, which converts $\{s_{k,n}^{(d)}\}$ to $\{f_{k,m}^{(d)}\}$. These are written in a $K_d \times N_c$ matrix $\hat{\bm{D}}^{(d)}$, whose rows and columns are for users and subcarriers, respectively. With $\bm{G}^{(d)} \triangleq \left[ \bm{g}_1^{(d)} \cdots \bm{g}_{K_d}^{(d)} \right]$ and $\bm{F}^{(d)}=\bm{T}\bm{D}^{(d)}$, it is expressed as
\begin{equation} \label{eq_hatDd}
    \hat{\bm{D}}^{(d)} = \left(\bm{G}^{(d)}\right)^{H} \bm{T} \bm{D}^{(d)} + \bm{N}^{(d)}. 
\end{equation}

\subsection{Multi-User Equalization and Precoding} \label{sec_ch_eq}
Multi-user equalization for the UL communication can be implemented via the least-squares (LS) method \cite{Kay93} to obtain the soft symbol estimates $\hat{\bm{D}}^{(u)}$ as 
\begin{equation} \label{eq_ULChEq}
    \hat{\bm{D}}^{(u)} = \left( (\hat{\bm{G}}^{(u)})^{H} \hat{\bm{G}}^{(u)} \right)^{-1} (\hat{\bm{G}}^{(u)})^{H} (\bm{Z}-\hat{\bm{Z}}_{si})  .
\end{equation}
In this procedure, the estimated UL channel $\hat{\bm{G}}^{(u)}$ and the regenerated SI signal $\hat{\bm{Z}}_{si}$ is used, to be detailed in the next sections. The obtained symbol estimates are expressed in terms of the actual ones as
\begin{equation} \label{eq_DhatU}
    \hat{\bm{D}}^{(u)} = \bm{M}^{(u)} \bm{D}^{(u)} + \bm{\Xi}^{(u)}
\end{equation}
where $\bm{\Xi}^{(u)}$ contains self-interference and the noise, and
\begin{equation} 
    \bm{M}^{(u)} \triangleq \left( (\hat{\bm{G}}^{(u)})^{H} \hat{\bm{G}}^{(u)} \right)^{-1} (\hat{\bm{G}}^{(u)})^{H} \bm{G}^{(u)}.
\end{equation}
For the UL user $k$, the symbol estimates can be written as 
\begin{equation} 
    \hat{\bm{d}}_k^{(u)} = \underbrace{m_{kk}^{(u)} \bm{d}_k^{(u)}}_{\text{intended symbols}} + \underbrace{\sum_{j=1,\, j \neq k}^{K_u}  m_{kj}^{(u)} \bm{d}_j^{(u)}}_{\text{multi-user interference}} + \underbrace{\bm{\xi}_k^{(u)}}_{\text{SI and noise}}
\end{equation}
where $m_{ij}^{(u)}\triangleq \left(\bm{M}^{(u)}\right)_{ij}$. In addition, $\hat{\bm{d}}_i^{(u)}$, $\bm{d}_i^{(u)}$, and $\bm{\xi}_i^{(u)}$ are the $i$\textsuperscript{th} row of $\hat{\bm{D}}^{(u)}$, $\bm{D}^{(u)}$, and $\bm{\Xi}^{(u)}$, respectively. Note that $\bm{M}^{(u)}$ is ideally the identity matrix and $m_{ij}^{(u)}=\delta[i-j]$ for the perfect estimation case where $\hat{\bm{G}}^{(u)} = \bm{G}^{(u)}$. So, the objective of the equalization is to mitigate the multi-user interference.

The corresponding objective at the DL side is called multi-user precoding, implemented by the BS through the precoder matrix $\bm{T}$. It is designed as 
\begin{equation}
    \bm{T} \triangleq \hat{\bm{G}}^{(d)}  \left( (\hat{\bm{G}}^{(d)})^{H} \hat{\bm{G}}^{(d)} \right)^{-1}
\end{equation}
which is called zero-forcing precoding. Using \eqref{eq_hatDd}, it yields
\begin{equation} 
    \hat{\bm{D}}^{(d)} = \bm{M}^{(d)} \bm{D}^{(d)} + \bm{N}^{(d)},
\end{equation}
where $\bm{M}^{(d)} \triangleq (\bm{G}^{(d)})^H \bm{T}$, which equals the identity matrix for the perfect channel estimation case where $\hat{\bm{G}}^{(d)}=\bm{G}^{(d)}$.

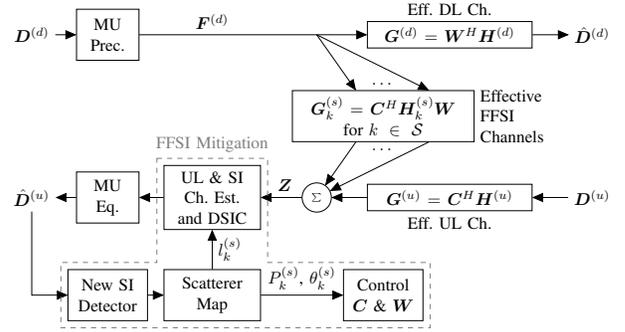
\begin{figure}[bt]
\centering
\resizebox{0.45\textwidth}{!}{%
\begin{tikzpicture}[>=triangle 45]

\def\sc{1.4}  
\def\ysc{1.2*\sc} 

\def\xdel{1.4*\sc}
\def\xa{0}
\def\xb{\xa + \xdel}
\def\xc{\xb + 0.8*\xdel}
\def\xd{\xc + 1.1*\xdel}
\def\xe{\xd + 1.1*\xdel}
\def\xf{\xe + 0.7*\xdel}
\def\xg{\xf + 0.7*\xdel}
\def\xh{\xg + 1.5*\xdel}

\def\ydel{1*\ysc}
\def\ya{0}
\def\yb{\ya - \ydel}
\def\yc{\yb - \ydel}
\def\yd{\yc - \ydel}
\def\ye{\yd - 1.2*\ydel}

\def\eps{0.1*\sc}

\node (dd) at (\xb,\yb) {$\bm{D}^{(d)}$};
\node (dzf) [block, text width=3em, minimum height=3em] at (\xc,\yb) {MU\\Prec.};
\node (fdtext) [anchor=south] at (\xd,\yb) {$\bm{F}^{(d)}$};
\node (dch) [block, text width=9em] at (\xg,\yb) {$\bm{G}^{(d)}=\bm{W}^H \bm{H}^{(d)}$};
\node (dchtext) [anchor=south] at (dch.north) {Eff. DL Ch.};
\node (dd_) at (\xh,\yb) {$\hat{\bm{D}}^{(d)}$};

\node (sch) [block, minimum height=3em, text width=10em] at (\xf,\yc) {$\bm{G}_k^{(s)}=\bm{C}^H \bm{H}_k^{(s)} \bm{W}$\\for $k \in \mathcal{S}$};
\node (schtext) [anchor=west, text width=5em] at (sch.east) {Effective\\FFSI\\Channels};

\node (ud_) at (\xb,\yd) {$\hat{\bm{D}}^{(u)}$};
\node (uzf) [block, minimum height=3em, text width=3em] at (\xc,\yd) {MU\\Eq.};
\node (udsic) [block, minimum height=3em, text width=5em] at (\xd,\yd) {UL \& SI Ch. Est. and DSIC};
\node (usum) [summer] at (\xe,\yd) {\tiny $\sum$};
\node (uch) [block, text width=9em] at (\xg,\yd) {$\bm{G}^{(u)}=\bm{C}^H \bm{H}^{(u)}$};
\node (uchtext) [anchor=north] at (uch.south) {Eff. UL Ch.};
\node (ud) at (\xh,\yd) {$\bm{D}^{(u)}$};

\node (nsid) [block, minimum height=3em, text width=4em] at (\xc,\ye) {New SI\\Detector};
\node (map) [block, minimum height=3em, text width=5em] at (\xd,\ye) {Scatterer\\Map};
\node (cw) [block, minimum height=3em, text width=4em] at (\xf,\ye) {Control\\$\bm{C}$ \& $\bm{W}$};
\node (cwtext) [anchor=south] at ($(map.east)!0.5!(cw.west)$) {$P_k^{(s)}$, $\theta_k^{(s)}$};
\node (dsictext) [anchor=west] at ($(map.north)!0.5!(udsic.south)$) {$l_k^{(s)}$};
\node (ztext) [anchor=south] at ($(usum.west)!0.5!(udsic.east)+(0.5*\eps,0)$) {$\bm{Z}$};

\draw[->] (dd) -- (dzf);
\draw[->] (dzf) -- (dch);
\draw[->] (dch) -- (dd_);
\draw[<-] (ud_) -- (uzf);
\draw[<-] (uzf) -- (udsic);
\draw[<-] (udsic) -- (usum);
\draw[<-] (usum) -- (uch);
\draw[<-] (uch) -- (ud);
\draw[->] (dzf-|usum) -- ($(sch.north west)!0.35!(sch.north east)$);
\draw[<-] (usum) -- ($(sch.south west)!0.35!(sch.south east)$);
\draw[->] (dzf-|usum) -- ($(sch.north west)!0.75!(sch.north east)$);
\draw[<-] (usum) -- ($(sch.south west)!0.75!(sch.south east)$);
\node () [anchor=south] at (sch.north) {$\dots$};
\node () [anchor=north] at (sch.south) {$\dots$};

\draw[->] (ud_) |- (nsid);
\draw[->] (nsid) -- (map);
\draw[->] (map) -- (udsic);
\draw[->] (map) -- (cw);

\node (uchtext) [anchor=south,gray] at ($(udsic.north)+(0,\eps)$) {FFSI Mitigation};

\draw[dashed,gray] ($(udsic.north west)+(-\eps,\eps)$) -- ($(udsic.north east)+(\eps,\eps)$) -- ($(udsic.south east)+(\eps,-\eps)$) -- ($(udsic.east|-cw.north)+(\eps,\eps)$) -- ($(cw.north east)+(\eps,\eps)$) -- ($(cw.south east)+(\eps,-\eps)$) -- ($(nsid.south west)+(-\eps,-\eps)$) -- ($(nsid.north west)+(-\eps,\eps)$) -- ($(nsid.north-|udsic.west)+(-\eps,\eps)$) -- ($(udsic.north west)+(-\eps,\eps)$);

\end{tikzpicture}
}
\caption{The equivalent system diagram to the general one in Fig. \ref{fig_sys_diag}. It considers the frequency domain without FFT-IFFT pairs and the beamformers inside the effective (Eff.) channels, and it focuses on the FFSI assuming NFSI mitigation is handled by well-studied methods in the literature. (MU: Multi-user, Prec.: Precoder, Eq.: Equalizer, Est.: Estimator).}
\label{fig_sys_diag_eq}
\end{figure}

\subsection{Power Expressions} \label{sec_power_exp}
For beamforming design in Section \ref{sec_spatsup} and the derivation of prior performance metrics in Section \ref{sec_selection}, we need to express powers of different terms in the observation $\bm{Z}$, which is the output of the Rx beamformer and input of the DSIC stage as seen in Figures \ref{fig_sys_diag} and \ref{fig_sys_diag_eq}. We can decompose the entry of $\bm{Z}$ for the $k_u$\textsuperscript{th} UL user in the $m$\textsuperscript{th} subcarrier as
\begin{equation}
    (\bm{Z})_{k_u,m} = \underbrace{z_{k_u,m}^{(is)}}_{\text{intended signal}} + \underbrace{z_{k_u,m}^{(mi)}}_{\text{multi-user int.}} + \underbrace{z_{k_u,m}^{(si)}}_{\text{SI}} + \underbrace{z_{k_u,m}^{(n)}}_{\text{noise}}.
\end{equation}
Using the definitions given in the introduced signal model, the powers of these terms are calculated as
\begin{align}
   \text{var}(z_{k_u,m}^{(is)}) & = P_{k_u}^{(u)} |\bm{c}_{k_u}^H \bm{q}_{k_u}^{(u)}|^2 ,
   \\
   \text{var}(z_{k_u,m}^{(mi)}) & = \sum_{k_u'=1, \, k_u' \neq k_u}^{K_u} P_{k_u'}^{(u)} |\bm{c}_{k_u}^H \bm{q}_{k_u'}^{(u)}|^2 ,
   \\
   \text{var}(z_{k_u,m}^{(si)}) & = \sum_{k_s=1}^{K_s} \sum_{k_d=1}^{K_d} P_{k_s}^{(s)} |\bm{w}_{k_d}^H \bm{q}_{k_s}^{(s)}|^2 |\bm{c}_{k_u}^H \bm{q}_{k_s}^{(s)}|^2  ,\label{eq_power_si}
   \\
   \text{var}(z_{k_u,m}^{(n)}) & = N_0 \bm{c}_{k_u}^H \bm{c}_{k_u},
\end{align}
where $\text{var}(\cdot)$ is the variance operator. For DL, the intended signal power is $P_{k_d}^{(d)} |\bm{w}_{k_d}^H \bm{q}_{k_d}^{(d)}|^2$ and the noise power is $N_0$.

In the signal model, transmit power and path loss are governed only by the channel coefficients $\alpha_k^{(x)}$ and their variances $P_k^{(x)}$ for $x\in \{u,d,s\}$ for simplicity. To relate to the physical parameters, we can write the UL signal power at a single element of BS Rx array: $P_{R,UL}$, and the SI power at a single Rx element due to a single Tx element: $P_{R,SI}$, as
\begin{align}
    P_{R,UL} & =P_{UE} \frac{\lambda^2}{(4\pi)^2 d_u^2},
& 
    P_{R,SI} & =\frac{P_{BS}}{N_a} \frac{\lambda^2 \sigma}{(4\pi)^3 d_s^4}, 
\end{align}
using Friis' transmission and radar range equations, respectively, for isotropic antenna elements with unity gains.\footnote{Antenna gains for whole arrays would depend on beamforming weights and user/scatterer positions. Instead, the element-to-element definitions of $P_{R,UL}$ and $P_{R,SI}$ overcome this dependency and provide a general validity.} In these equations, $\lambda$ is the wavelength, $P_{UE}$ is the user Tx power, $P_{BS}$ is total Tx power for BS, $\sigma$ is radar cross section (RCS) for scatterers, and $d_u$ and $d_s$ are the distances of the user and scatterer, respectively. In addition, we can also write $P_{R,UL}$ and $P_{R,SI}$ according to our signal model as
\begin{align} \label{eq_P_R_UL}
    P_{R,UL} & = \mathbb{E} \{(\bm{h}_k^{(u)} s_{k,n}^{(u)} )_{a_r}\} = \frac{P_k^{(u)}}{N_a},
\\
    P_{R,SI} & = \mathbb{E} \{(\bm{H}_k^{(s)})_{a_r,a_t} 
    (\bm{W} \bm{s}_{n-l_k^{(s)}}^{(d)} )_{a_t}\} =\frac{P_k^{(s)}K_d}{N_a^3}, \label{eq_P_R_SI}
\end{align}
where $a_r$ and $a_t$ are arbitrary Rx and Tx antenna element indices, respectively. By equating the related expressions, the variances in our signal model are found as
\begin{align} \label{eq_ch_variances}
    P_k^{(u)} & = P_{UE} \frac{N_a \lambda^2}{(4\pi)^2 d_u^2}, & 
    P_k^{(s)} & = P_{BS} \frac{N_a^2\lambda^2 \sigma}{K_d (4\pi)^3 d_s^4},
\end{align}
which include the Tx powers, path loss, and scattering effects.

\section{SI Mitigation Methods}
The SI component in the observation $\bm{Z}$ might degrade the communication performance. To mitigate its effects, we will investigate \emph{DSIC} and \emph{spatial suppression} via transmit beamforming (Tx BF) and receive beamforming (Rx BF).

We consider a system that can apply these methods concurrently. 
Accordingly, for each scatterer, a method is selected among Rx BF, Tx BF, and DSIC. Let $\mathcal{S}_{Rx}$, $\mathcal{S}_{Tx}$, $\mathcal{S}_{DSIC}$ be the sets of scatterers for these methods, respectively. We consider also that no action is taken for some scatterers, forming $\mathcal{S}_{NA}$. These sets disjointly partition the set of all scatterers $\mathcal{S}$, mathematically expressed as
\begin{align}
    \bigcup_{a\in \mathcal{A}} \mathcal{S}_a & = \mathcal{S},
    \\
    \mathcal{S}_a \cap \mathcal{S}_{a'} & = \O, \qquad a, a' \in \mathcal{A}, \quad a \neq a',
\end{align}
where $\mathcal{A} \triangleq \{Tx, Rx, DSIC, NA\}$ is the set of actions. 

\subsection{Spatial SI Suppression} \label{sec_spatsup}
This approach aims to decrease the exposed SI power by optimizing the beamformers $\bm{C}$ and $\bm{W}$ statistically based on scatterer angles $\theta_k^{(s)}$. Note that this is mathematically equivalent to decreasing the Frobenius norm of SI channel $G_k^{(s)}$ defined in \eqref{eq_Gks}.  

With the sets $\mathcal{S}_{Rx}$ and $\mathcal{S}_{Tx}$, the well-known maximum signal-to-interference-plus-noise ratio (SINR) beamformer is used for Rx BF and the maximum signal-to-leakage-plus-noise ratio (SLNR) beamformer \cite{Sadek07} can be used for the Tx BF, as derived in the Appendix. They are written as
\begin{align}
    \label{eq_opt_c}
    \bm{c}_k= \beta_k^{(c)} & \hspace{-2pt} \left(N_0 \bm{I}_{N_a} \hspace{-2pt} + \hspace{-7pt} \sum_{k' \in \mathcal{S}_{Rx}} \hspace{-4pt} \rho_{k'}^{(s,c)} \bm{q}_{k'}^{(s)} (\bm{q}_{k'}^{(s)})^{H}\hspace{-2pt}\right)^{-1}  \hspace{-8pt}\bm{q}_{k}^{(u)} 
    \\
    \label{eq_opt_w}
    \bm{w}_k= \beta_k^{(w)} & \hspace{-2pt} \left(\dfrac{K_u}{K_d}N_0 \bm{I}_{N_a} \hspace{-2pt} + \hspace{-7pt} \sum_{k' \in \mathcal{S}_{Tx}} \hspace{-4pt} \rho_{k'}^{(s,w)} \bm{q}_{k'}^{(s)} (\bm{q}_{k'}^{(s)})^{H}\hspace{-2pt}\right)^{-1}  \hspace{-8pt}\bm{q}_{k}^{(d)}
\end{align}
where $\beta_k^{(c)}$ and $\beta_k^{(w)}$ normalizes norms of $\bm{c}_k$ and $\bm{w}_k$ to unity, respectively. In addition, $\rho_{k'}^{(s,c)} \triangleq P_{k'}^{(s)} \sum_{k''} |(\bm{q}_{k''}^{(d)})^H \bm{q}_{k'}^{(s)}|^2$ and $\rho_{k'}^{(s,w)} \triangleq P_{k'}^{(s)} \sum_{k''} |(\bm{q}_{k''}^{(u)})^H \bm{q}_{k'}^{(s)}|^2$.%
\footnote{Differently from the Appendix, note that $\bm{c}_k=\bm{q}_k^{(u)}$ is assumed for $\rho_{k}^{(s,w)}$, and $\bm{w}_k=\bm{q}_k^{(d)}$ is assumed for $\rho_{k}^{(s,c)}$. This approximation breaks the dependency between $\bm{c}_k$ and $\bm{w}_k$ to avoid iterations by sacrificing a slight improvement. Actually, the beamformer on one side can reduce the SI power from a scatterer, which relaxes the other beamformer. However, this would be negligible since $\mathcal{S}_{Rx} \cap \mathcal{S}_{Tx} = \O$.} 
Since $\bm{c}_k$ is the $k$\textsuperscript{th} column of $\bm{C}$, the matrix can be written as
\begin{equation}
    \bm{C} =  \left(N_0 \bm{I} + \bm{Q}_{\mathcal{S}_{Rx}} \bm{P}_{\mathcal{S}_{Rx}} \bm{Q}_{\mathcal{S}_{Rx}}^{H}\right)^{-1} \bm{Q}^{(u)} \bm{B}^{(c)}
\end{equation}
where $\bm{Q}^{(u)}$ and $\bm{Q}_{\mathcal{S}_{Rx}}$ contain $\bm{q}_{k}^{(u)}$ for $k \in \mathcal{U}_u$ and $\bm{q}_{k}^{(s)}$ for $k \in \mathcal{S}_{Rx}$ in their columns, respectively. In addition, the diagonal matrices $\bm{P}_{\mathcal{S}_{Rx}}$ and $\bm{B}^{(c)}$ contain $\rho_{k}^{(s,c)}$ for $k \in \mathcal{S}_{Rx}$ and $\beta_k^{(c)}$ for $k \in \mathcal{U}_u$ in their diagonal, respectively. Using Woodburry's identity, the Rx beamformer $\bm{C}$ can be expressed in the HBF structure as 
\begin{equation} \label{eq_hyb_bf_structure}
    \bm{C} =  \frac{1}{N_0} \underbrace{\begin{bmatrix}[\bm{Q}^{(u)}] & [\bm{Q}_{\mathcal{S}_{Rx}}] \end{bmatrix}}_{\text{ABF: } N_a \times (K_u+|\mathcal{S}_{Rx}|)} \underbrace{\begin{bmatrix} [\bm{C}_D^{(u)}]^{T} & [\bm{C}_D^{(s)}]^{T} \end{bmatrix}^{T}}_{\text{DBF: } (K_u+|\mathcal{S}_{Rx}|) \times K_u},
\end{equation}
with analog and digital beamformers (ABF and DBF). ABF columns have constant modulus and progressive phase. The parts in DBF are defined as $\bm{C}_D^{(u)}\triangleq \bm{B}^{(c)}$ and
\begin{equation}
    \bm{C}_D^{(s)} \triangleq - \left(N_0 \bm{P}_{\mathcal{S}_{Rx}}^{-1} + \bm{Q}_{\mathcal{S}_{Rx}}^{H} \bm{Q}_{\mathcal{S}_{Rx}}\right)^{-1}\bm{Q}_{\mathcal{S}_{Rx}}^{H}\bm{Q}^{(u)} \bm{B}^{(c)}.
\end{equation}
As seen, the proposed Rx beamformer can also be realized in the HBF structure. This applies to the Tx beamformer too because the expressions are similar. Therefore, SI mitigation actions Rx BF and Tx BF require $K_u+|\mathcal{S}_{Rx}|$ and $K_d+|\mathcal{S}_{Tx}|$ RFCs, respectively.

\subsection{Digital SI Cancellation (DSIC)} \label{sec_dsic}
Using the designed beamformers, the UL and DL communication is performed. In the UL communication, the frequency domain observation matrix $\bm{Z}$ is obtained both for the data and the training. Using the training on $N_{c,tr}\ll N_c$ subcarriers,%
\footnote{We consider that the $N_{c,tr}$ training subcarriers are uniformly chosen among $N_c$ subcarriers. This training/data distinction should be indicated on variables such as $\bm{Z}$, $\bm{D}^{(u)}$, $\bm{D}^{(d)}$, $\bm{E}_l$ and many more, which would result in a complex notation. We skip these indications for notational simplicity since the training/data distinction is clear from the context.}
the UL channel is estimated jointly with the SI channels for the scatterers inside $\mathcal{S}_{DSIC}$. Let the indices in the set $\mathcal{S}_{DSIC}$ be denoted by $k_i^{DSIC}$. The joint LS estimator for the related SI and SoI channels is
\begin{equation} \label{eq_dsic_chest}
    \left[ \hat{\bm{G}}^{(u)} \hat{\bm{G}}_{k_1^{DSIC}}^{(s)} \hat{\bm{G}}_{k_2^{DSIC}}^{(s)} \dots \right]
     = \bm{Z} \bar{\bm{D}}^{H} (\bar{\bm{D}} \bar{\bm{D}}^{H})^{-1},
\end{equation}
where related training symbol matrices are stacked vertically in the $(K_u+K_d |\mathcal{S}_{DSIC}|)\times N_{c,tr}$ matrix $\bar{\bm{D}}$ as 
\begin{equation}
\bar{\bm{D}} = \left[ \left[\bm{D}^{(u)}\right]^T \, \left[\bm{F}^{(d)} \bm{E}_{l_1}\right]^T \, \left[\bm{F}^{(d)} \bm{E}_{l_2}\right]^T \dots \, \right]^T,
\end{equation}
where $l_i$ is the delay for the scatterer $k_i^{DSIC}$. The SI channels are used for SI regeneration in the data mode as
\begin{equation} \label{eq_dsic_regen}
    \hat{\bm{Z}}_{si} = \sum_{k \in \mathcal{S}_{DSIC}} \hat{\bm{G}}_k^{(s)} \bm{F}^{(d)} \bm{E}_{l_k^{(s)}},
\end{equation}
to be subtracted from the received signal as $\bm{Z}-\hat{\bm{Z}}_{si}$. After this SIC procedure, the UL channel equalization is performed to obtain estimates for the UL symbols as in \eqref{eq_ULChEq}.

\section{Selection Policy for SI Mitigation} \label{sec_selection}
This section aims to determine a method for each scatterer to construct the sets $\mathcal{S}_{Rx}$, $\mathcal{S}_{Tx}$, $\mathcal{S}_{DSIC}$, and $\mathcal{S}_{NA}$. Each scatterer is handled with the best action that maximizes the communication performance based on the prior performance metrics, which are approximations of signal-to-noise ratio (SNR) for DL and SINR for UL. Deriving these, each scatterer is considered separately with the assumed signal model
\begin{equation} \label{eq_Z_for_metrics}
    \bm{Z} = \bm{G}^{(u)} \bm{D}^{(u)} +  \bm{G}_{k_s}^{(s)} \bm{F}^{(d)} \bm{E}_{l_{k_s}^{(s)}} + \bm{N}
\end{equation}
for the scatterer $k_s$, where other scatterers in \eqref{eq_Z} are neglected. Although we will express $\bm{c}_k$ and $\bm{w}_k$ based on \eqref{eq_Z_for_metrics} in derivation, beamformer implementations remain as described in \eqref{eq_opt_c} and \eqref{eq_opt_w}. Since the scatterer $k_s$ will be focused, we drop this index from variables $\bm{G}_{k_s}^{(s)}$, $\alpha_{k_s}^{(s)}$, $P_{k_s}^{(s)}$, $l_{k_s}^{(s)}$ and $\bm{q}_{k_s}^{(s)}$ until \eqref{eq_mean_metric} for simplicity.

The power expressions in Section \ref{sec_power_exp} can be used directly to write the UL SINR $\gamma_k^{(u)}$ and DL SNR $\gamma_k^{(d)}$, modifying only the SI power in \eqref{eq_power_si} for a single scatterer. In addition, the multi-user interference is ignored in the metrics because it will be handled by the multi-user equalizer after FFSI mitigation, as shown in Fig. \ref{fig_sys_diag_eq}. Accordingly, the metrics are written as
\begin{align}
    \gamma_k^{(u)} & = \frac{P_k^{(u)} |\bm{c}_k^H \bm{q}_{k}^{(u)}|^2} {P^{(s)} \sum_{k'} |\bm{w}_{k'}^H \bm{q}^{(s)}|^2 |\bm{c}_k^H \bm{q}^{(s)}|^2 + N_0}, && k \in \mathcal{U}_u
    \label{eq_general_sinr}
    \\
    \gamma_k^{(d)} & = \frac{P_k^{(d)} |\bm{w}_k^H \bm{q}_{k}^{(d)}|^2} {N_0}, && k \in \mathcal{U}_d \label{eq_general_snr}
\end{align}
where $\bm{c}_k^H\bm{c}_k=1$ and $\bm{w}_k^H\bm{w}_k=1$. In the next sections, $\gamma_k^{(u)}$ and $\gamma_k^{(d)}$ will be written for each case $a \in \mathcal{A}$, denoted by $\gamma_k^{(u,a)}$ and $\gamma_k^{(d,a)}$. Note that these metrics are for the scatterer $k_s$ in parallel to \eqref{eq_Z_for_metrics}, and they will be regenerated for each scatterer in $\mathcal{S}$ in Section \ref{sec_algorithm}.

\subsection{No-SI and No-Action Cases}
In the no-SI and no-action cases, the Rx and Tx beamformer weights are conventionally taken as $\bm{c}_k=\bm{q}_{k}^{(u)}$ and $\bm{w}_k=\bm{q}_{k}^{(d)}$. 
For the no-SI case, the UL SINR and DL SNR are
\begin{align}
    \gamma_k^{(u,0)} = \frac{P_k^{(u)}} { N_0 }, &&   \gamma_k^{(d,0)} = \frac{P_k^{(d)}} { N_0 },
\end{align}
respectively. The metrics for the no-action case are
\begin{align} \label{eq_metric_UL_NA}
    \gamma_k^{(u,NA)} = \frac{\gamma_k^{(u,0)}} {1 + \frac{P^{(s)}}{N_0} \sum_{k'=1}^{K_d} Q_{k'}^{(ds)} Q_k^{(us)}},
\end{align}
and $\gamma_k^{(d,NA)} = \gamma_k^{(d,0)}$, where $Q_k^{(ds)}$ and $Q_k^{(us)}$ govern the effects of user/scatterer positioning, defined as 
\begin{align} \label{eq_Q}
Q_k^{(ds)} \triangleq |(\bm{q}_k^{(d)})^H \bm{q}^{(s)}|^2,
&& 
Q_k^{(us)} \triangleq |(\bm{q}_k^{(u)})^H \bm{q}^{(s)}|^2.
\end{align}

\subsection{Max-SINR Rx Beamformer}
In this case, the Tx beamformer is $\bm{w}_k= \bm{q}_{k}^{(d)}$, and the Rx beamformer in \eqref{eq_opt_c} is written for a single scatterer as 

\begin{equation} \label{eq_first_exp_of_c_metric}
    \bm{c}_k= \beta_k^{(c)}\left(N_0 \bm{I} + \rho_s \bm{q}^{(s)} (\bm{q}^{(s)})^{H}\right)^{-1} \bm{q}_{k}^{(u)},
\end{equation}
where $\rho_s \triangleq P^{(s)} \sum_{k'} |(\bm{q}_{k'}^{(d)})^H \bm{q}^{(s)}|^2$. Then, the DL SNR is simply
$\gamma_k^{(d,Rx)} = \gamma_k^{(d,0)}$
from \eqref{eq_general_snr}, and the UL SINR can be calculated from \eqref{eq_general_sinr}. For a closed-form expression, we can rewrite the Rx BF using Woodburry's identity as
\begin{align}
    \bm{c}_k & = \beta_k^{'(c)} \left(\bm{q}_{k}^{(u)} - \frac{(\bm{q}^{(s)})^{H} \bm{q}_{k}^{(u)}}{1 + N_0/\rho_s} \bm{q}^{(s)} \right),
    \\
     & \cong \beta_k^{'(c)} \left(\bm{q}_{k}^{(u)} - [(\bm{q}^{(s)})^{H} \bm{q}_{k}^{(u)}] \bm{q}^{(s)} \right), \label{eq_RxMet_c_HighINR}
\end{align}
where $\beta_k^{'(c)}\triangleq \frac{\beta_k^{(c)}}{N_0}$, and \eqref{eq_RxMet_c_HighINR} implements a high INR assumption ($\rho_s/N_0 \rightarrow \infty$). Then, $||\bm{c}_k||^2=1$ is satisfied by $\beta_k^{'(c)} = (1 - Q_k^{(us)})^{-\frac{1}{2}}$, which yields
\begin{align}
    |\bm{c}_k^H \bm{q}_{k}^{(u)}|^2 & = 1 - Q_k^{(us)},
    \\
    |\bm{c}_k^H \bm{q}^{(s)}|^2 & = 0,
\end{align}
for $\gamma_k^{(u,Rx)}$ in \eqref{eq_general_sinr}. Therefore, the metrics are calculated as
\begin{align}
    \gamma_k^{(u,Rx)} = \gamma_k^{(u,0)} (1 - Q_k^{(us)}), && \gamma_k^{(d,Rx)} = \gamma_k^{(d,0)}
\end{align}

\subsection{Max-SLNR Tx Beamformer}
In this case, the Rx beamformer is 
$\bm{c}_k=\bm{q}_{k}^{(u)}$, and the Tx beamformer in \eqref{eq_opt_w} is modified as 
\begin{equation} \label{eq_first_exp_of_w_metric}
    \bm{w}_k= \beta_k^{(w)} \left(\dfrac{K_u}{K_d} N_0 \bm{I} + \rho_s \bm{q}^{(s)} (\bm{q}^{(s)})^{H}\right)^{-1} \bm{q}_{k}^{(d)},
\end{equation}
where $\rho_s \triangleq P^{(s)} \sum_{k'} |(\bm{q}_{k'}^{(u)})^H \bm{q}^{(s)}|^2$. 
Note that the expression in \eqref{eq_first_exp_of_w_metric} is in the same form as \eqref{eq_first_exp_of_c_metric} except the scaling $N_0$ of the identity matrix, which did not affect the obtained metrics. Hence, following similar steps, the metrics can be found as
\begin{align}
    &\gamma_k^{(u,Tx)} = \gamma_k^{(u,0)}, & \gamma_k^{(d,Tx)} = \gamma_k^{(d,0)} (1 - Q_k^{(ds)}).
\end{align}
\subsection{Digital SI Cancellation (DSIC)}
We assume no spatial effort for DSIC with conventional beamformers 
$\bm{c}_k=\bm{q}_{k}^{(u)}$ and $\bm{w}_k=\bm{q}_{k}^{(d)}$, which directly sets the DL SNR as $\gamma_k^{(d,DSIC)} = \gamma_k^{(d,0)}$. Then, the assumed form of $\bm{Z}$ in \eqref{eq_Z_for_metrics} can also be written as
\begin{equation} 
    \bm{Z} = \left[\bm{G}^{(u)} \bm{G}^{(s)} \right]  \bar{\bm{D}} + \bm{N}
\end{equation}
where $\bar{\bm{D}} = \left[[\bm{D}^{(u)}]^T \quad [\bm{F}^{(d)} \bm{E}_l]^T\right]^T$. The LS estimates for channel matrices are given by $\bm{Z} \bar{\bm{D}}^{H} (\bar{\bm{D}} \bar{\bm{D}}^{H})^{-1}$, which yields
\begin{equation}
    \left[ \hat{\bm{G}}^{(u)} \hat{\bm{G}}^{(s)} \right]
    = \left[\bm{G}^{(u)} \bm{G}^{(s)} \right] + \bm{N} \bar{\bm{D}}^{H} (\bar{\bm{D}} \bar{\bm{D}}^{H})^{-1}
\end{equation}
The matrix $\bar{\bm{D}}$ is of size $(K_u+K_d)\times N_{c,tr}$, which carries training sequences for UL and DL users, each of length $N_{c,tr}$. Each symbol is unit-power and the sequences are linearly independent. With the further assumption that they are orthogonal, $\bar{\bm{D}} \bar{\bm{D}}^{H}$ becomes $N_{c,tr} \bm{I}_{K_u+K_d}$. Then, the estimates are rewritten as
\begin{equation}
    \left[ \hat{\bm{G}}^{(u)} \hat{\bm{G}}^{(s)} \right]
     =  \left[\bm{G}^{(u)} \bm{G}^{(s)} \right] + \frac{1}{N_{c,tr}} \bm{N} \bar{\bm{D}}^{H}.
\end{equation}
from which the SI channel estimation error is found as
\begin{equation}
\bm{\Delta}_s \triangleq \hat{\bm{G}}^{(s)} - \bm{G}^{(s)} = \frac{1}{N_{c,tr}} \bm{N} \left[\bm{F}^{(d)} \bm{E}_l \right]^{H}.
\end{equation}
Columns of $\bm{N}$ are independent and identically distributed (i.i.d.) with $\mathcal{CN}(\bm{0},N_0 \bm{C}^{H} \bm{C})$. Since DL training symbols are assumed orthogonal, the error matrix $\bm{\Delta}_s$ has i.i.d. columns and each column follows $\mathcal{CN}(\bm{0},\frac{N_0}{N_{c,tr}} \bm{C}^{H} \bm{C})$. When SI is regenerated as $\hat{\bm{Z}}_{si} = \hat{\bm{G}}^{(s)}\bm{F}^{(d)} \bm{E}_l$ and subtracted, we obtain
\begin{equation}
    \bm{Z} - \hat{\bm{Z}}_{si} = \bm{G}^{(u)} \bm{D}^{(u)} - \bm{\Delta}_s \bm{F}^{(d)} \bm{E}_l + \bm{N},
\end{equation}
which is different from \eqref{eq_Z} only in the SI term. Power expressions for \eqref{eq_Z} were written in Section \ref{sec_power_exp}, so we can use those expressions for the intended signal and the noise. The new term, the residual SI with power $P_{SI}^{DSIC}$, can be expressed for the UL user $k_u$ and the subcarrier $m$ as
\begin{equation}
    (\bm{\Delta}_s \bm{F}^{(d)} \bm{E}_l)_{k_u,m} = \sum_{k_d=1}^{K_d} (\bm{\Delta}_s)_{k_u,k_d} \, f_{k_d,m}^{(d)} \, e^{-j \frac{2\pi}{N_{c,tr}} m l}.
\end{equation}
Then, the variance of this term gives $P_{SI}^{DSIC}$ as 
\begin{align}
   P_{SI}^{DSIC} = & \sum_{k_d} \text{var}\left((\bm{\Delta}_s)_{k_u,k_d}\right) |f_{k_d,m}^{(d)} e^{-j \frac{2\pi}{N_{c,tr}} m l}|^2 \label{eq_P_SI_DSIC_1}
   \\ = & \sum_{k_d} \frac{N_0}{N_{c,tr}} \bm{c}_{k_u}^H \bm{c}_{k_u} |f_{k_d,m}^{(d)}|^2 = \frac{N_0}{N_{c,tr}} \sum_{k_d} |f_{k_d,m}^{(d)}|^2 
   \\ = & \frac{K_d N_0}{N_{c,tr}}, \label{eq_P_SI_DSIC_3}
\end{align}
where we used the independence of columns of $\bm{\Delta}_s$ in \eqref{eq_P_SI_DSIC_1} and the unit average symbol energy property in \eqref{eq_P_SI_DSIC_3}. Finally, the DL SNR is written similarly to \eqref{eq_metric_UL_NA} as
\begin{equation}
    \gamma_k^{(u,DSIC)} = \frac{\gamma_k^{(u,0)}} { 1 +  \frac{K_d}{N_{c,tr}} }
\end{equation}
which differs only in the SI power.
\subsection{Determining Actions for Scatterers Based on the Metrics} \label{sec_algorithm}

\begin{table}[b]
\centering
\caption{Prior Performance Metrics}
\label{tab_metrics}
\begin{tabular}{|c|c|c|}
\hline
\multirow{2}*{\textbf{Cases} ($a$)} & \textbf{Normalized UL SINR} & \textbf{Norm. DL SNR} 
\\ 
& $\gamma_{k_u}^{(u,a)}/\gamma_{k_u}^{(u,0)}$ & $\gamma_{k_d}^{(d,a)}/\gamma_{k_d}^{(d,0)}$
\\[0.5ex] \hline \hline
\rule{0pt}{11pt} 
No Action ($NA$) & \hspace{-5pt} $ {\scriptstyle \left(1 + \frac{P^{(s)}}{N_0} \sum_{k_d} Q_{k_d}^{(ds)} Q_{k_u}^{(us)} \right)^{-1}}$ \hspace{-5pt} & $1$
\\[1ex] \hline 
\rule{0pt}{10pt} 
Rx BF ($Rx$) & $1 - Q_{k_u}^{(us)}$ & $1$
\\[0.5ex] \hline
\rule{0pt}{10pt} 
Tx BF ($Tx$) & $1$ & $1 - Q_{k_d}^{(ds)}$
\\[0.5ex] \hline 
\rule{0pt}{9pt} 
DSIC ($DSIC$) & $\left(1 + {K_d/N_{c,tr}}\right)^{-1}$ & $1$
\\[0.3ex] \hline 
\end{tabular}
\end{table}

The approximated UL SINRs and DL SNRs are tabulated in Table \ref{tab_metrics}. When focused on a single scatterer $k_s \in \mathcal{S}$ and a single action $a \in \mathcal{A}$, we have SINR or SNR for each user, such as $\gamma_1^{(u,a)},\dots, \gamma_{K_u}^{(u,a)},\gamma_1^{(d,a)},\dots, \gamma_{K_d}^{(d,a)}$. Their arithmetic mean gives a single metric for this scatterer and action pair as
\begin{equation} \label{eq_mean_metric}
    \bar{\gamma}_{k_s}^{(a)} = \frac{1}{K_u+K_d} \left(\sum_{k=1}^{K_u} \frac{\gamma_k^{(u,a)}}{\gamma_k^{(u,0)}}+\sum_{k=1}^{K_d} \frac{\gamma_k^{(d,a)}}{\gamma_k^{(d,0)}}\right)
\end{equation}
where we reinstate the scatterer index $k_s$ that was dropped after \eqref{eq_Z_for_metrics} for simplicity. 

The average performance metric $\bar{\gamma}_{k_s}^{(a)}$ is calculated for each scatterer/action pair. Among these metrics, $\bar{\gamma}_{k_s}^{(NA)}$ tells the severity of the scatterer $k_s$, and the others tell the improvement expected from Rx BF, Tx BF, and DSIC. If there are no limitations, each scatterer should be dealt with the best action. 

\subsubsection{Limited Resource Case (Hybrid Beamformer)} 
To cover also the case of limited resources, e.g. RFCs for Rx BF and Tx BF, we introduce the predefined limits $L_a$ as
\begin{equation}
    |\mathcal{S}_a| \le L_a
\end{equation}
for $a \in \mathcal{A}$. Note that $L_a=K_s$ for each $a \in \mathcal{A}$ for the default case of no limitations, and $\sum_{a\in\mathcal{A}} L_a \ge K_s$ should always be satisfied. Also, we will only deal with the limited RFC case that is common for massive MIMO devices, for which $L_{NA}=L_{DSIC}=K_s$, $L_{Tx}< K_s$ and $L_{Rx}< K_s$. Under these limitations, prioritizing the most severe scatterers according to $\bar{\gamma}_{k_s}^{(NA)}$, each scatterer is associated with the best available option. In the end, sets of scatterers are obtained for each action, namely $\mathcal{S}_{NA}$, $\mathcal{S}_{Rx}$, $\mathcal{S}_{Tx}$, and $\mathcal{S}_{DSIC}$. This procedure is detailed in Algorithm \ref{algorithm}.

\begin{algorithm}[bt]
\begin{algorithmic}[1]
\REQUIRE $\theta_{k_u}^{(u)}$, $\theta_{k_d}^{(d)}$, $\theta_{k_s}^{(s)}$, $P_{k_s}^{(s)}$, $L_a$ for $k_u \in \mathcal{U}_u$, $k_d \in \mathcal{U}_d$, $k_s \in \mathcal{S}$, $a \in \mathcal{A}$.

\FOR{$k_s=1$ to $K_s$}
\STATE set $\bm{q}^{(s)}=\bm{q}_{k_s}^{(s)}$ and $\bm{P}^{(s)}=\bm{P}_{k_s}^{(s)}$.
\STATE calculate $Q_{k_u}^{(us)}$, $Q_{k_d}^{(ds)}$ for $k_u\in \mathcal{U}_u$, $k_d\in \mathcal{U}_d$. (Eq. \eqref{eq_Q})
\STATE calculate $\bar{\gamma}_{k_s}^{(a)}$ for $a \in \mathcal{A}$. (Eq. \eqref{eq_mean_metric} and Table \ref{tab_metrics})
\ENDFOR
\STATE $\mathcal{S}_{a}=\O$ for $a \in \mathcal{A}$, $\mathcal{S}'=\mathcal{S}$, and $\mathcal{A}'=\{ a \in \mathcal{A} \, : \, L_a > 0 \}$.
\WHILE{$|\mathcal{S}'|>0$}
\STATE $k_s=\underset{k_s'\in \mathcal{S}'}{\argmin} \,  \bar{\gamma}_{k_s'}^{(NA)}$
\STATE $a^* = \underset{a \in \mathcal{A}'}{\argmax} \, \bar{\gamma}_{k_s}^{(a)}$
\STATE $\mathcal{S}_{a^*} \leftarrow  \mathcal{S}_{a^*} \cup \{k_s\}$
\STATE $\mathcal{S}' \leftarrow  \mathcal{S}' - \{k_s\}$
\IF {$|\mathcal{S}_{a^*}|=L_{a^*}$} 
\STATE $\mathcal{A}' \leftarrow  \mathcal{A}' - \{a^*\}$ \ENDIF
\ENDWHILE
\ENSURE $\mathcal{S}_{a}$ for $a \in \mathcal{A}$.
\end{algorithmic}
    \caption{Construction of Action Sets.}
    \label{algorithm}
\end{algorithm}

\begin{figure}[bt]
\centering
\resizebox{0.45\textwidth}{!}{%
\begin{tikzpicture}[>=triangle 45]

\def\sc{0.9}  
\def\ysc{1.8*\sc} 

\def\xdel{1.4*\sc}
\def\xa{0}
\def\xb{\xa + 0.4*\xdel}
\def\xc{\xb + 0.4*\xdel}
\def\xd{\xc + 1.2*\xdel}
\def\xe{\xd + 1.0*\xdel}
\def\xf{\xe + 1.0*\xdel}
\def\xg{\xf + 1.0*\xdel}
\def\xh{\xg + 1.0*\xdel}
\def\xi{\xh + 1.0*\xdel}
\def\xj{\xi + 1.2*\xdel}
\def\xk{\xj + 0.4*\xdel}
\def\xl{\xk + 0.4*\xdel}

\def\xcd{\xc + 0.7*\xdel}
\def\xfga{\xf + 0.2*\xdel}
\def\xfgb{\xf + 0.8*\xdel}
\def\xhi{\xh + 0.4*\xdel}

\def\ydel{1*\ysc}
\def\ya{0}
\def\yb{\ya - 0.8*\ydel}
\def\yc{\yb - 0.7*\ydel}
\def\yd{\yc - 0.6*\ydel}
\def\ye{\yd - 0.75*\ydel}
\def\yf{\ye - 0.7*\ydel}
\def\yg{\yf - 0.8*\ydel}

\def\eps{0.1*\sc}

\draw (\xc,\yb) -- (\xc,\yc);
\draw (\xd,\yb) -- (\xd,\yc);
\draw[dashed] (\xe,\yb) -- (\xe,\yc);
\draw[dashed] (\xf,\yb) -- (\xf,\yc);
\draw[dashed] (\xg,\yb) -- (\xg,\yc);
\draw[dashed] (\xh,\yb) -- (\xh,\yc);
\draw (\xi,\yb) -- (\xi,\yc);
\draw (\xj,\yb) -- (\xj,\yc);

\draw[dashed] (\xa,\yb) -- (\xb,\yb);
\draw (\xb,\yb) -- (\xfga,\yb);
\draw[dashed] (\xfga,\yb) -- (\xfgb,\yb);
\draw (\xfgb,\yb) -- (\xk,\yb);
\draw[dashed] (\xk,\yb) -- (\xl,\yb);
\draw[dashed] (\xa,\yc) -- (\xb,\yc);
\draw (\xb,\yc) -- (\xfga,\yc);
\draw[dashed] (\xfga,\yc) -- (\xfgb,\yc);
\draw (\xfgb,\yc) -- (\xk,\yc);
\draw[dashed] (\xk,\yc) -- (\xl,\yc);
\draw [fill={rgb:black,2;white,8}] (\xc,\yb) rectangle (\xd,\yc);
\draw [fill={rgb:black,2;white,8}] (\xi,\yb) rectangle (\xj,\yc);

\node () [text centered, text width=5em] at ($(\xc,\yb)!0.5!(\xd,\yc)$) {Sensing \& Tracking};
\node () [fill=white, text centered] at ($(\xd,\yb)!0.5!(\xi,\yc)$) {Full-Duplex Communication};
\node () [text centered, text width=5em] at ($(\xi,\yb)!0.5!(\xj,\yc)$) {Sensing \& Tracking};

\node (asaf1) [text centered, text width=10em] at (\xd,\ya) {Action sets are fixed.\\Rx \& Tx BF are set.};
\draw[->] (asaf1) -- (\xd,\yb);
\node (new) [text centered, text width=8em] at (\xg,\ya) {New scatterer emerges.};
\draw[->] (new) -- (\xg,\yb);
\node (asaf2) [text centered, text width=10em] at (\xj,\ya) {Action sets are fixed.\\Rx \& Tx BF are set.};
\draw[->] (asaf2) -- (\xj,\yb);

\node (rawdata) [block, rounded corners, text centered, text width=5em] at (\xb,\yd) {Raw Data In};
\node (chest) [block, text centered, text width=5em] at (\xcd,\ye) {UL \& SI Ch. Est.};
\node (dsic) [block, text centered, text width=5em] at (\xf,\ye) {DSIC \& Ch. Eq.};
\node (newsi) [block, text centered, text width=5em, rounded corners] at (\xhi,\ye) {New SI Detector};
\node (dataout) [block, rounded corners, text centered, text width=5em] at (\xl,\yd) {Processed Data Out};
\node (dsicsetupdate) [block, rounded corners, text width=15em, align=left, anchor=north east] at ({(\xh,\yf)}-|dataout.east) {$\cdot$ $\mathcal{S} \leftarrow \mathcal{S} \cup \{k^*\}$\\$\cdot$ $\mathcal{S}_{DSIC} \leftarrow \mathcal{S}_{DSIC} \cup \{k^*\}$\\$\cdot$ Repeat from "UL \& SI Ch. Est."\\$\cdot$ Operate "Initial Angle Est."};
\node (angleout) [anchor=south east] at ($(dsicsetupdate.west)-(6pt,0)$)  {$\hat{\theta}_{k^*}^{(s)}$};
\node (angleest) [block, text centered, text width=5em, anchor=east] at ($(angleout.west)-(10pt,0)$) {Initial Angle Est.};
\node (newchanin) [anchor=east] at ($(angleest.west)-(10pt,0)$) {$\hat{\bm{G}}_{k^*}^{(s)}$};

\node (si0) [anchor=south west] at (newsi.10) {No new SI};
\node (si1) [anchor=north west] at (newsi.-10) {New SI at $\hat{l}_{k^*}^{(s)}$};

\draw[->] (rawdata) |- (chest);
\draw[->] (chest) -- (dsic);
\draw[->] (dsic) -- (newsi);
\draw[->] (newsi.10) -| (dataout);
\draw[->, dashed] (newsi.-10) -- (newsi.-10-|dataout) -- (dataout|-dsicsetupdate.north);
\draw[->, dashed] (dsicsetupdate.195) -| (chest);
\draw[->] (newchanin) -- (angleest);
\draw[->] (angleest) -- (angleout);
\draw[->, dashed] (chest.-10) -| ($(chest.south east)!0.5!(dsic.south west)$) |- ($(newchanin.north)!0.5!(chest.south-|newchanin.north)$) -- (newchanin);

\node (sym_datain) [anchor=south east] at (rawdata|-chest) {$\bm{Z}$};
\node (sym_dataout) [anchor=south west] at (newsi.10-|dataout) {$\hat{\bm{D}}^{(u)}$};
\node (sym_chest) [anchor=south, text width=5em, text centered] at ($(chest)!0.5!(dsic)$) {$\hat{\bm{G}}^{(u)}$, $\{\hat{\bm{G}}_k^{(s)}\}$};
\node (sym_dataeqd) [anchor=south] at ($(dsic)!0.5!(newsi)$) {$\hat{\bm{D}}^{(u)}$};

\draw (\xe,\yc) -- (\xe,\yd);
\draw (\xf,\yc) -- (\xf,\yd);
\draw (\xg,\yc) -- (\xg,\yd);
\draw (\xh,\yc) -- (\xh,\yd);
\draw (\xi,\yc) -- (\xi,\yd);
\draw (\xi,\yd) -- (\xfgb,\yd);
\draw[dashed] (\xfgb,\yd) -- (\xfga,\yd);
\draw[->] (\xfga,\yd) -- (rawdata);
\end{tikzpicture}
}
\caption{Regular communication and sensing intervals, and processing flow for a possible emerging scatterer during communication.}
\label{fig_jsac}
\end{figure}
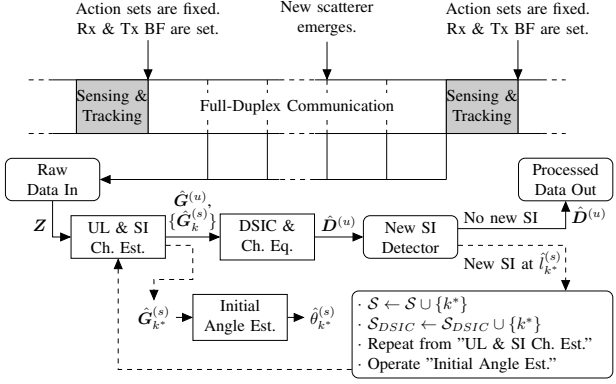

\section{Scatterer Map} \label{sec_EST_ind}
The proposed SI mitigation scheme, which consists of spatial SI suppression and DSIC, is applied using some slow-time parameters regarding the scatterers. These are the delay $l_k^{(s)}$, the angle $\theta_k^{(s)}$, and the power $P_k^{(s)}$, stored in the \emph{scatterer map} shown in Fig. \ref{fig_sys_diag_eq}. Among these, DSIC uses the delay, and spatial suppression uses the angle and the power. 

\subsection{Sensing and Tracking}
We consider sensing and tracking phases as in Fig. \ref{fig_jsac} to keep the scatterer map up-to-date, which is a time-shared form of ISAC. After this interval, we assume that the delay $l_k^{(s)}$, angle $\theta_k^{(s)}$, and power $P_k^{(s)}$ are available to determine the action sets. Although we consider a separate sensing and tracking interval, useful information can be collected during communication too. 

The probable changes in the scatterer map are emergence, disappearance, and variations due to mobility. While emerging scatterers should be sensed, others can be tracked using the available information. Our system model enables tracking during communication to some extent with constantly estimated SI channels in DSIC, as shown in \eqref{eq_dsic_chest}. Also, extra analog beams are directed towards the scatterers in Rx BF for better suppression, as shown in \eqref{eq_hyb_bf_structure}. While the DBF in \eqref{eq_hyb_bf_structure} suppresses SI, another parallel DBF structure after the available ABF may deliberately observe it. Although these are not the main focus of this paper, emerging scatterers during communication might be problematic for FD communications. 

\subsection{Processing for an Emerging Scatterer} \label{sec_emerging}
The emergence of an unknown scatterer during the FD communication interval, as shown in Fig. \ref{fig_jsac}, might cause a loss of information. Detection of this new scatterer might prevent this by triggering an extra mitigation effort, which can be considered as \emph{sensing for communication}. Besides detection and mitigation, the system can also estimate some prior information about the scatterer, e.g. the angle, which might help in the upcoming sensing stage. That is, the search objective might be replaced by verification and refinement. This section will detail the proposed detection, mitigation, and initial angle estimation flow illustrated in Fig. \ref{fig_jsac}.

To formulate the problem, we define hypotheses $H_1(l)$ and $H_0(l)$ for the presence and absence of an emerging SI at the delay $l$, respectively. After the SI mitigation, we can rewrite the symbol estimates in \eqref{eq_DhatU} for these cases as
\begin{align} \label{eq_no_new_si}
    H_0(l): \quad & \hat{\bm{D}}^{(u)} = \bm{D}^{(u)} + \tilde{\bm{N}},
    \\
    H_1(l): \quad & \hat{\bm{D}}^{(u)} = \bm{D}^{(u)} + \bm{M}^{(s)} \bm{F}^{(d)} \bm{E}_{l^*} + \tilde{\bm{N}},\label{eq_with_new_si}
\end{align}
where $\tilde{\bm{N}} \triangleq (\bm{M}^{(u)}-\bm{I})\bm{D}^{(u)} + \bm{\Xi}^{(u)}$ contains the term due to UL channel estimation error besides residual SI from known scatterers and the thermal noise. In addition, $k^*$ and $l^*$ are the index and delay of the emerging scatterer, respectively, and $\bm{M}^{(s)}$ is the modified version of the new SI channel $\bm{G}_{k^*}^{(s)}$ seen after the multi-user equalizer, which is written as
\begin{equation}
\bm{M}^{(s)} \triangleq \left( (\hat{\bm{G}}^{(u)})^{H} \hat{\bm{G}}^{(u)} \right)^{-1} (\hat{\bm{G}}^{(u)})^{H} \bm{G}_{k^*}^{(s)}.
\end{equation} 

\subsubsection{Detection and Delay Estimation}
Observing hypotheses \eqref{eq_no_new_si} and \eqref{eq_with_new_si}, a detector can be designed on an estimate of $\bm{M}^{(s)}$. It seems suitable to process  $\hat{\bm{D}}^{(u)} - \bm{D}^{(u)}$ for this estimate. The $r$\textsuperscript{th} row of that matrix is written as
\begin{equation}
    (\hat{\bm{D}}^{(u)}-\bm{D}^{(u)})_{r,:} = \sum_{t=1}^{K_d} (\bm{M}^{(s)})_{r,t} (\bm{F}^{(d)})_{t,:} \bm{E}_l 
    + (\tilde{\bm{N}})_{r,:}
\end{equation}
Among the estimators investigated in \cite{Kurt_pimrc} for similar purposes, it is seen that the suboptimal \emph{per-user SI channel estimator} performs similarly with the other complex estimators in detection, although it fails in SI regeneration. This simple estimator is adapted for our purposes since complexity might be an issue for this task. It estimates the entries of $\bm{M}^{(s)}$ separately as
\begin{equation}
    \left( \hat{\bm{M}}_l^{(s)} \right)_{r,t} = 
    \frac{1}{\sqrt{N_{c,tr}}} \text{IFFT} \{ 
    (\hat{\bm{D}}^{(u)}-\bm{D}^{(u)})_{r,:} \oslash (\bm{F}^{(d)})_{t,:}
    \}_l,
\end{equation}
for $r=1,\dots,K_u$ and $t=1,\dots,K_d$, where $\oslash$ is elementwise division. Candidate estimates $\hat{\bm{M}}_l^{(s)}$ are obtained for each delay $l \in \mathcal{L}$, where $\mathcal{L}$ is the set of possible delays. Then, Frobenius norm squares of these matrices are tested as
\begin{align}
    &\rho_l \underset{H_0(l)}{\overset{H_1(l)}{\gtrless}}   \alpha, &&\rho_l \triangleq ||\hat{\bm{M}}_l^{(s)}||_F^2,
\end{align}
which decides $H_1(l)$ or $H_0(l)$ for each $l \in \mathcal{L}$. We determine the threshold $\alpha$ by the constant false alarm rate (CFAR) method \cite{book_richards}, which fixes the false alarm probability $P_{FA} \triangleq P(\rho_l>\alpha \, |H_0(l))$ to a desired value by adjusting the $\alpha$. However, the cumulative distribution function (CDF) of $\rho_l$ given $H_0(l)$ is needed, which is hard to obtain due to the complex structure of $\tilde{\bm{N}}$. 

Alternatively, a \emph{cell-averaging} CFAR test \cite{book_richards} requires only the type of the distribution, where the unknown variance is empirically calculated from the metric $\rho_l$ for $l \in \mathcal{L}$. Assuming each entry of $\hat{\bm{M}}_l^{(s)}$ is i.i.d. and follows $\mathcal{CN}(0,\sigma^2)$ given $H_0(l)$ with an unknown $\sigma^2$, the test is written as \cite{Kurt_pimrc}
\begin{equation} \label{eq_det}
    \rho_l \underset{H_0(l)}{\overset{H_1(l)}{\gtrless}}   \frac{g(P_{FA})}{2 K_u K_d} \left( \frac{1}{(|\mathcal{L}|-1)} \sum_{l' \in \mathcal{L}, l'\neq l} \rho_{l'} \right),
\end{equation}
where $g(P_{FA})$ sets the CDF of $\chi^2$ distribution with $2 K_u K_d$ degrees of freedom to $1-P_{FA}$. The scalings convert the mean of this distribution, $2 K_u K_d$, to the empirical mean. 

\subsubsection{Repetition of SI Channel Estimation and DSIC}
In the case of detection, the system immediately mitigates the related SI by DSIC, since spatial suppression may not be viable\footnote{With HBFs, the spatial suppression requires changes in the analog beamformer, which is impossible once it is set and reception is completed. With fully digital beamformers, spatial suppression is possible but it requires storage of the full-dimensional received signal.}. The detected scatterer at delay $l^*$ is provided with an index $k^*=K_s+1$, which sets $l_{k^*}^{(s)}=l^*$. Then, $k^*$ is added to the set of scatterers for DSIC as $\mathcal{S}_{DSIC}\leftarrow \mathcal{S}_{DSIC} \cup \{k^*\}$. The channel estimation and DSIC operations are repeated with the new set as in Section \ref{sec_dsic}. With this repetition, besides having the new SI channel estimate $\hat{\bm{G}}_{k^*}^{(s)}$, previously estimated channels of users and scatterers are refined. Therefore, the multi-user channel equalization is also repeated as in Section \ref{sec_ch_eq} with the new UL channel estimate.

\subsubsection{Initial Angle Estimation} \label{sec_EST_angle}
Having the SI channel estimate $\hat{\bm{G}}_{k^*}^{(s)}$ for the emerging scatterer from the previous step, its angle can be estimated too. Using the general definition in \eqref{eq_Gks}, we can rewrite the SI channel as
\begin{equation}
    \text{vec}\{\bm{G}_k^{(s)}\} = \alpha_k^{(s)} \bm{v}(\theta_k^{(s)}),
\end{equation}
where $\text{vec}\{\cdot\}$ is the vectorization operator and $\bm{v}(\theta) \triangleq \text{vec}\{\bm{C}^{H} \bm{a}(\theta) \bm{a}^{H}(\theta) \bm{W}\}$. Using the estimate $\hat{\bm{G}}_{k^*}^{(s)}$, a \emph{nonlinear LS} estimator\footnote{This estimator from \cite{Kay93} combines LS estimate for $\alpha_k^{(s)}$ and maximum likelihood (ML) estimator for $\theta_k^{(s)}$, since both are unknown. The ML metric is $|| \text{vec}\{\hat{\bm{G}}_{k^*}^{(s)}\} - \alpha_k^{(s)} \bm{v}(\theta)\} ||^2$ assuming uncorrelated and Gaussian estimation errors. Then, the LS estimate for $\alpha_k^{(s)}$ is found for tested $\theta$ and substituted.}
for $\theta_{k^*}^{(s)}$ is written as \cite{Kurt_pimrc}
\begin{equation} \label{eq_ang_est}
    \hat{\theta}_{k^*}^{(s)} = \text{arg}\min_\theta || \text{vec}\{\hat{\bm{G}}_{k^*}^{(s)}\} - \frac{\bm{v}(\theta) \bm{v}^{H}(\theta)}{\bm{v}^{H}(\theta) \bm{v}(\theta)} \text{vec}\{\hat{\bm{G}}_{k^*}^{(s)}\} ||^2.
\end{equation}

This angular estimate can be directly used, or it can be taken as primary information to be verified or refined. In this way, the scatterer map is updated for the new scatterer with its delay $l_{k^*}^{(s)}$ and angle $\theta_{k^*}^{(s)}$. Using these, the power $P_{k^*}^{(s)}$ can also be estimated through a longer observation. Note that the detector in \eqref{eq_det} can detect disappearance and the estimator in \eqref{eq_ang_est} can refine the angle, as tracking efforts on known scatterers. 

\section{Complexity Analysis}
The required number of multiplications for various operations is presented in Table \ref{tab_complexity} using the $\mathcal{O}$ notation, where generally $N_c \gg N_{c,tr} > |\mathcal{S}_a|\cong K_x$. A key difference in complexity between DSIC and spatial suppression is that DSIC requires frequent operations, whereas spatial suppression leverages angles that remain valid for longer durations.

DSIC needs a modified channel estimation as shown in \eqref{eq_dsic_chest}, increasing the complexity by $\mathcal{O}\left(K_d^2 |\mathcal{S}_{DSIC}|^2 N_{c,tr}\right)$. However, the SI regeneration in \eqref{eq_dsic_regen} is the main burden with $\mathcal{O}\left(K_u K_d |\mathcal{S}_{DSIC}| N_c\right)$. Additionally, the constraint $N_{c,tr} > (K_u + K_d |\mathcal{S}_{DSIC}|)$ for invertibility adds to the complexity.

As discussed in Section \ref{sec_spatsup}, the cost of spatial SI suppression is the requirement for the extra RFCs, which are $|\mathcal{S}_{Tx}|$ for Tx BF and $|\mathcal{S}_{Rx}|$ for Rx BF. The extra computational complexity due to spatial SI suppression comes only from digital beamforming. Beamformer weight calculations can be disregarded, as these operations are infrequent compared to other processing tasks, as shown in Fig. \ref{fig_jsac}. This also applies to determining the actions based on metrics, described in Algorithm \ref{algorithm}. 

DSIC and spatial suppression need some resources, which might be limited for practical systems. While the exclusive use of a single method may be impossible under these limitations, the proposed selection policy offers a way to compromise the resources while maximizing performance. Algorithm \ref{algorithm} considers possible limitations inclusively, denoted by $L_a$. Although this work primarily investigates the case of limited RFCs, the proposed method is also applicable when the DSIC-related computational complexity should be limited.

\begin{table}[b]
    \centering
    \caption{Complexities of Operations}
    \label{tab_complexity}
    \begin{tabular}{|c|l|}
        \hline
        \textbf{Operation} & \# \textbf{Multiplications}\\
        \hline \hline
        \rule{0pt}{7pt} 
        Tx DBF & $\mathcal{O}\left( K_d (K_d + |\mathcal{S}_{Tx}|) N_c \right)$ \\
        \hline
        \rule{0pt}{7pt} 
        Rx DBF & $\mathcal{O}\left( K_u (K_u + |\mathcal{S}_{Rx}|) N_c \right)$ \\
        \hline
        \rule{0pt}{7pt} 
        FFT/IFFT & $\mathcal{O}\left( K_u N_c \log N_c \right)$ \\
        \hline
        \rule{0pt}{8pt} 
        Channel Estimation & $\mathcal{O}\left( (K_u + K_d |\mathcal{S}_{DSIC}|)^2 N_{c,tr} \right)$ \\
        \hline
        \rule{0pt}{7pt} 
        DSIC & $\mathcal{O}\left( K_u K_d |\mathcal{S}_{DSIC}| N_c \right)$ \\
        \hline
        \rule{0pt}{8pt} 
        Multi-user Equalization & $\mathcal{O}\left( K_u^2 N_c \right)$ \\
        \hline
        \rule{0pt}{10pt} 
        New SI Detector & $\mathcal{O}\left( K_u K_d N_{c,tr}^2 \log N_{c,tr}\right)$ \\
        \hline
    \end{tabular}
\end{table}
\section{Performance Evaluation}

For the simulations, considered system parameters are $N_a=32$ antenna elements both in Tx and Rx arrays of BS, $N_{c,tr}=64$ subcarriers for training, $P_k^{(x)}/N_0=10$ dB input SNR for $x\in\{u,d\}$, and $P_k^{(s)}/N_0=34$ dB INR.%
\footnote{$P_{R,UL}$ in \eqref{eq_P_R_UL} and $P_{R,SI}$ in \eqref{eq_P_R_SI} are equal for $P_{BS}/P_{UE}=20$ dB, $d_u=80$ m, $d_s=20$ m, and $\sigma = 100$ m\textsuperscript{2} RCS (typical for automobiles). This setting gives $P_k^{(s)}/P_k^{(u)}=24$ dB according to \eqref{eq_ch_variances}.} 
The numbers and positions of UL and DL users will be given while the figures are discussed. A pair of UL and DL users at the same location can be interpreted as a single FD user.  

In the simulations, the UL SINR $\Gamma_k^{(u)}$ and DL SNR $\Gamma_k^{(d)}$ are measured separately for each user as
\begin{align}
    \Gamma_k^{(x)} & = \frac{ || m_{kk}^{(x)} \bm{d}_k^{(x)} ||^2 }{ || \hat{\bm{d}}_k^{(x)} - m_{kk}^{(x)} \bm{d}_k^{(x)} ||^2 },
\end{align}
for $x \in \{u,d\}$. Then, simulations are repeated many times to obtain the performance metrics below.
\subsubsection{Worst-case UL SINR and DL SNR}
Worst-case UL SINR $\Gamma^{(u,w)}$ and worst-case DL SNR $\Gamma^{(d,w)}$ show the worst performance among multiple UL and DL users, shown as
\begin{equation}
    \Gamma^{(u,w)} = \min_{k\in\mathcal{U}_u} \mathbb{E}\{\Gamma_k^{(u)}\}, \qquad \Gamma^{(d,w)} = \min_{k\in\mathcal{U}_d} \mathbb{E}\{\Gamma_k^{(d)}\}.
\end{equation}

\subsubsection{AIR}
The achievable information rate (AIR) is 
\begin{equation}
    I_k^{(u)} = \mathbb{E}\left\{
    \log_2 \left(1 + \Gamma_k^{(u)} \right) 
     \right\}, \quad 
    I_k^{(d)} = \mathbb{E}\left\{
    \log_2 \left(1 + \Gamma_k^{(d)} \right) 
     \right\}
\end{equation}

\begin{figure}[tb]
\centering
\resizebox{0.45\textwidth}{!}{%
	\includegraphics{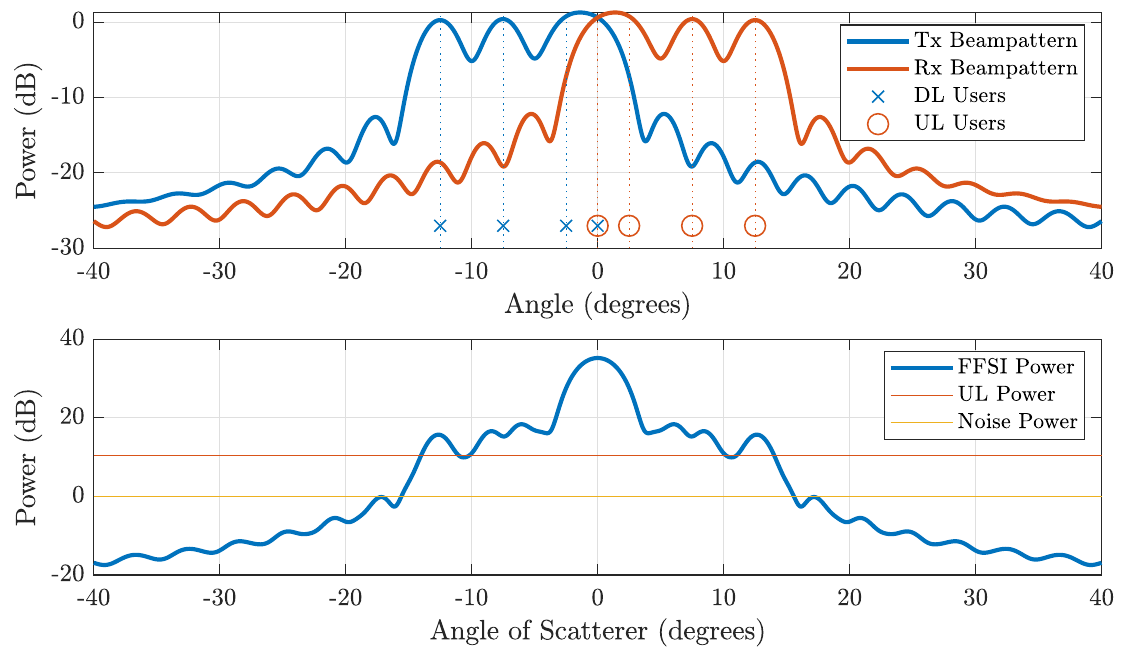}
	}
	\caption{Beampatterns and exposed far-field SI power. ($K_d=4$, $K_u=4$, $K_s=1$. User positions are as shown with markers.)}
	\label{fig_1}
\end{figure}

\begin{figure}[tb]
\centering
\resizebox{0.45\textwidth}{!}{%
	\includegraphics{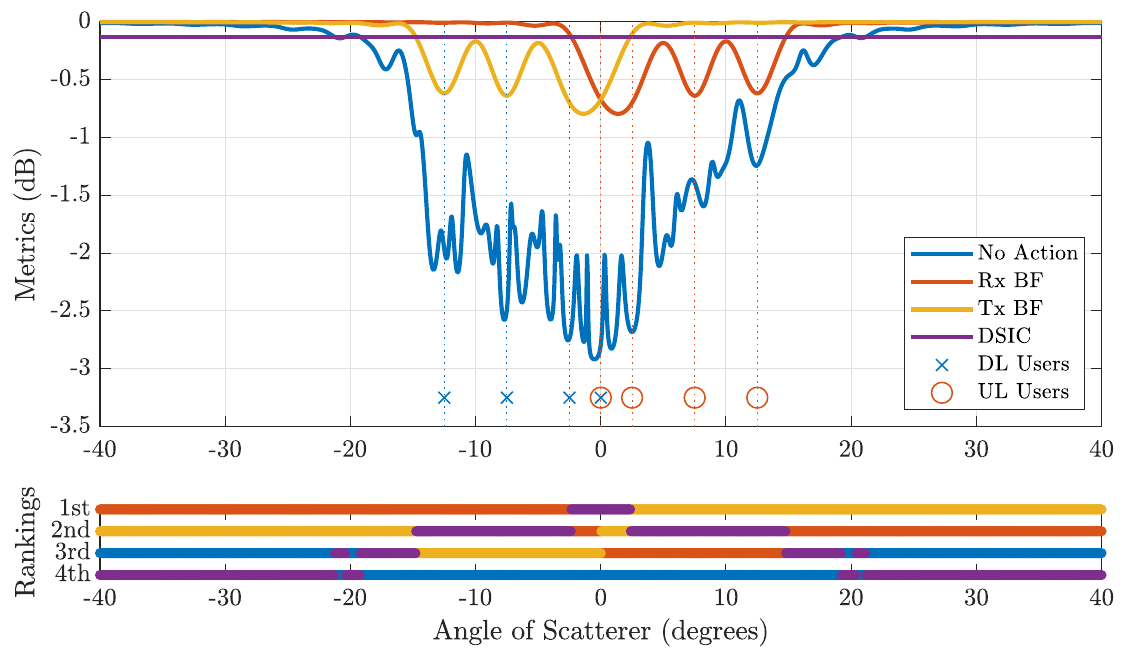}
	}
	\caption{Calculated action metrics and color-coded rankings of actions according to the angle of the considered scatterer. ($K_d=4$, $K_u=4$, $K_s=1$. User positions are as shown with markers.)}
	\label{fig_2}
\end{figure}

\begin{figure}[tb]
\centering
\resizebox{0.45\textwidth}{!}{%
	\includegraphics{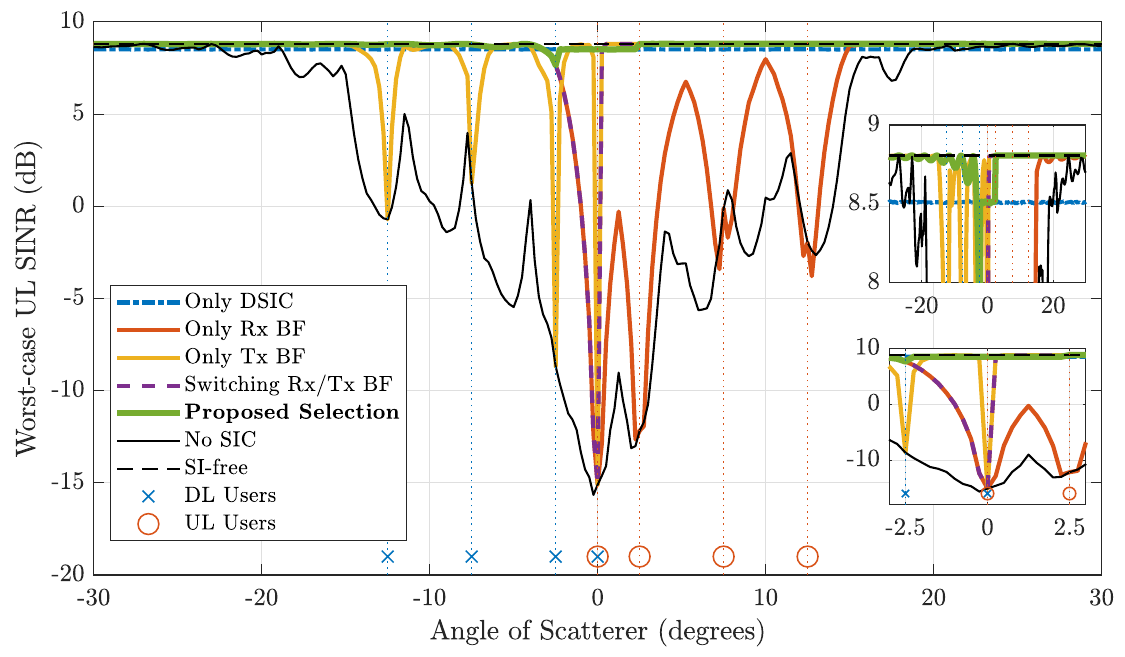}
	}
	\caption{Worst-case UL SINR against the angle of a single scatterer. ($K_d=4$, $K_u=4$, $K_s=1$. User positions are as shown with markers.)}
	\label{fig_3}
\end{figure}

\begin{figure}[tb]
\centering
\resizebox{0.45\textwidth}{!}{%
	\includegraphics{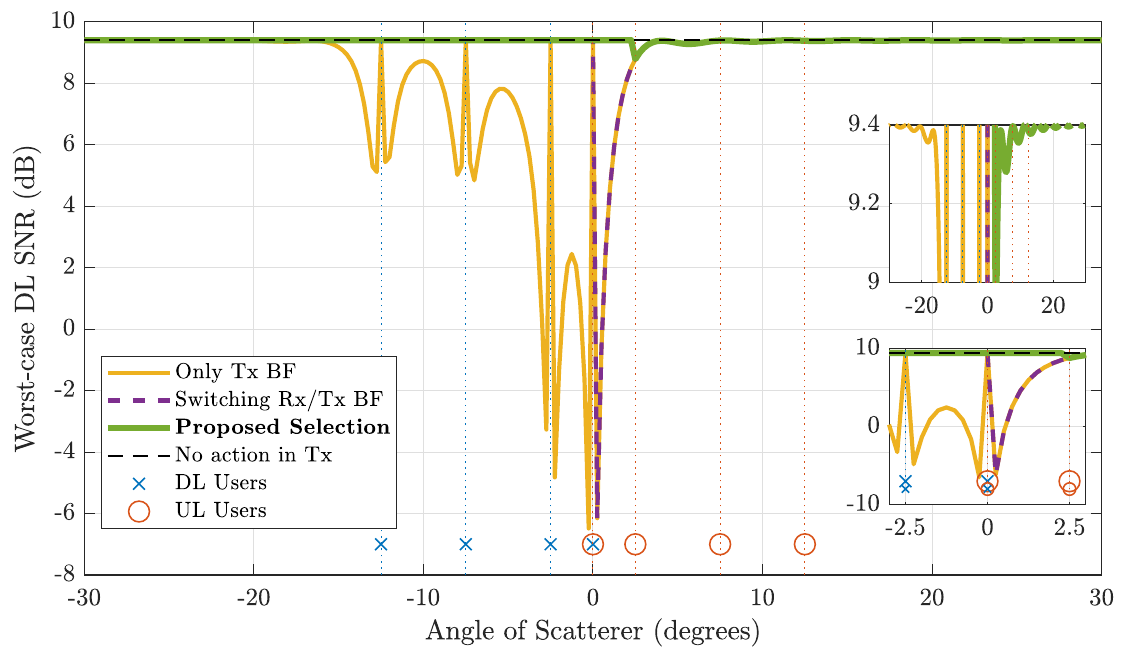}
	}
	\caption{Worst-case DL SNR against the angle of a single scatterer. ($K_d=4$, $K_u=4$, $K_s=1$. User positions are as shown with markers.)}
	\label{fig_4}
\end{figure}

\begin{figure}[tb]
\centering
\resizebox{0.45\textwidth}{!}{%
	\includegraphics{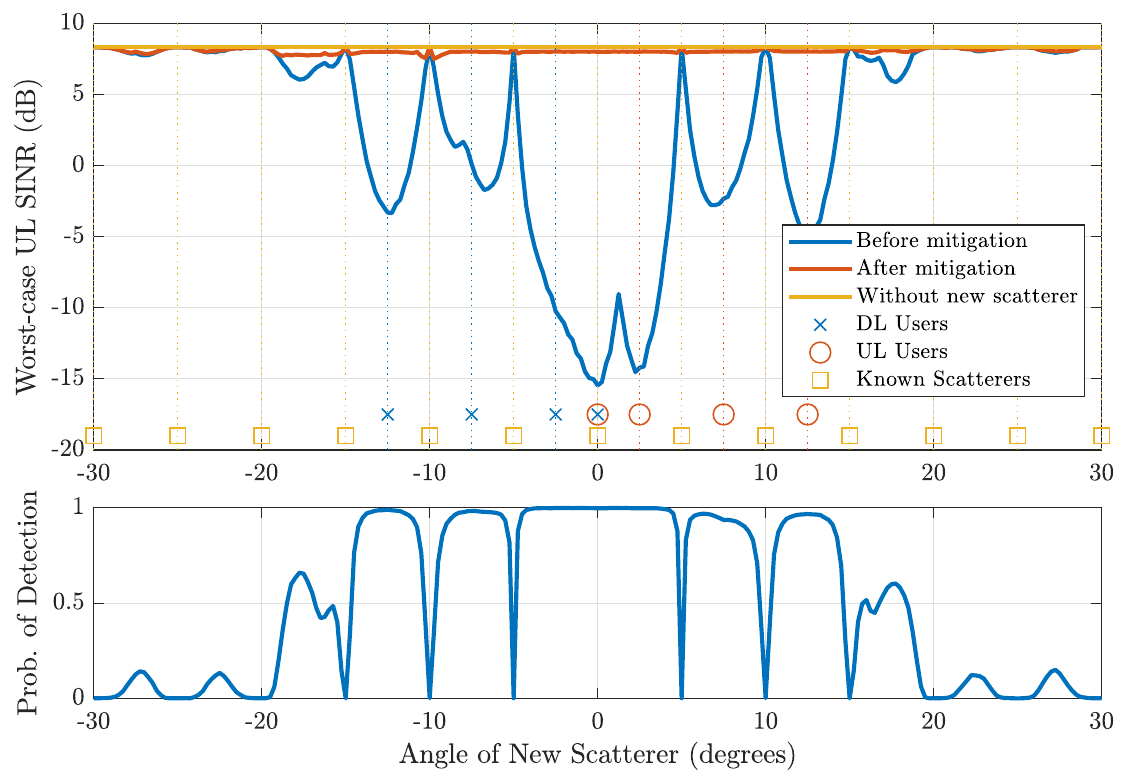}
	}
	\caption{Performance of detection and mitigation mechanisms in case of an emerging single scatterer. ($K_d=4$, $K_u=4$, $K_s=13$ for known scatterers. User and scatterer positions are as shown with markers.)}
	\label{fig_10}
\end{figure}

\begin{figure}[tb]
\centering
\resizebox{0.45\textwidth}{!}{%
	\includegraphics{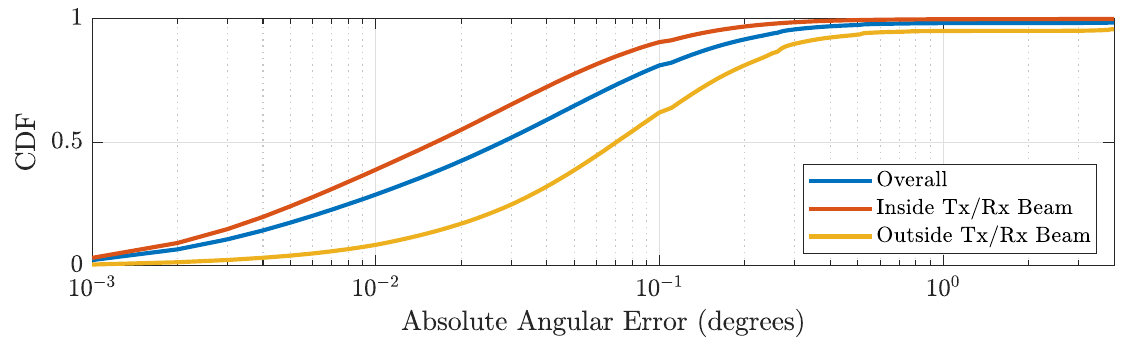}
	}
	\caption{Initial angular estimation accuracy in case of an emerging single scatterer. ($K_d=4$, $K_u=4$, $K_s=13$ for known scatterers. User and scatterer positions are as shown with markers in Fig. \ref{fig_10}.)}
	\label{fig_11}
\end{figure}

The beampatterns and exposed far-field SI power are plotted in Fig. \ref{fig_1}. As seen, due to the combined effect of Tx and Rx beampatterns, angular sectors are formed where the exposed SI power is higher than the received UL power, between the noise and the UL power, or lower than the noise power. In addition, Fig. \ref{fig_2} depicts the calculated action metrics according to the angle of the considered scatterer. For Fig. \ref{fig_2}, and also for Figures \ref{fig_3}, \ref{fig_4}, \ref{fig_10}, \ref{fig_11}, user angles are $\theta_k^{(u)}\in \{ 0^\circ , 2.5^\circ, 7.5^\circ, 12.5^\circ\}$ and $\theta_k^{(d)}\in \{ 0^\circ , -2.5^\circ, -7.5^\circ, -12.5^\circ\}$. Generally, it is seen that the scatterers near UL users require Tx BF, and those near DL users require Rx BF. When a scatterer is close to both UL and DL users, or an FD user, spatial suppression fails and DSIC is required. Finally, for a scatterer that is far away from the users, taking no action has negligible disadvantages.

Figures \ref{fig_3} and \ref{fig_4} show performances of different methods depending on the angle of a single scatterer. Rx BF deteriorates the UL performance when the scatterer is close to the UL users. Tx BF deteriorates both the UL and DL performance when the scatterer is close to DL users. To address this, we can follow a strategy where a scatterer is assigned to Tx BF when the closest user is UL and Rx BF when a DL user is closest. This strategy, labeled "Switching Rx/Tx BF" in the legends, fails when the scatterer is close to both a UL and a DL user, or a single FD user. On the other hand, DSIC seems very successful for this single scatterer case. As expected, DSIC performance does not depend on the scatterer angle but will be shown to depend on the number of scatterers. Finally, the proposed selection is the best and closest to the SI-free case, where the most suitable method is chosen based on the calculated metrics. Note that for this case, the calculated metrics can be explicitly seen from Fig. \ref{fig_2}, and the most suitable method can be followed from the "1st" row in the subplot below.

\begin{figure*}[tb]
\centering
\resizebox{0.90\textwidth}{!}{%
	\includegraphics{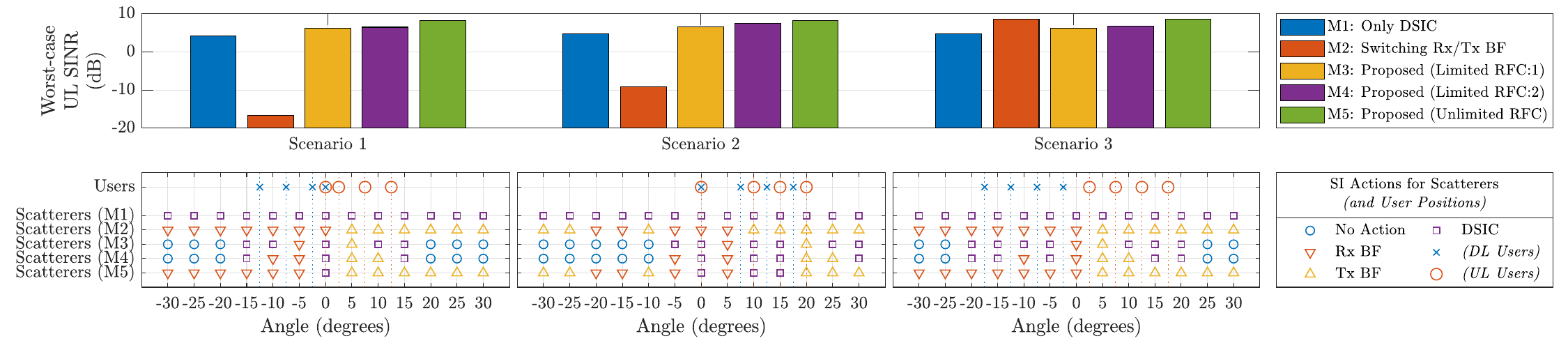}
	}
	\caption{Action choices and performances on exemplary scenarios. ($K_d=4$, $K_u=4$, $K_s=13$. User and scatterer positions are as shown with markers.)}
	\label{fig_9}
\end{figure*}

\begin{figure}[tb]
\centering
\resizebox{0.45\textwidth}{!}{%
	\includegraphics{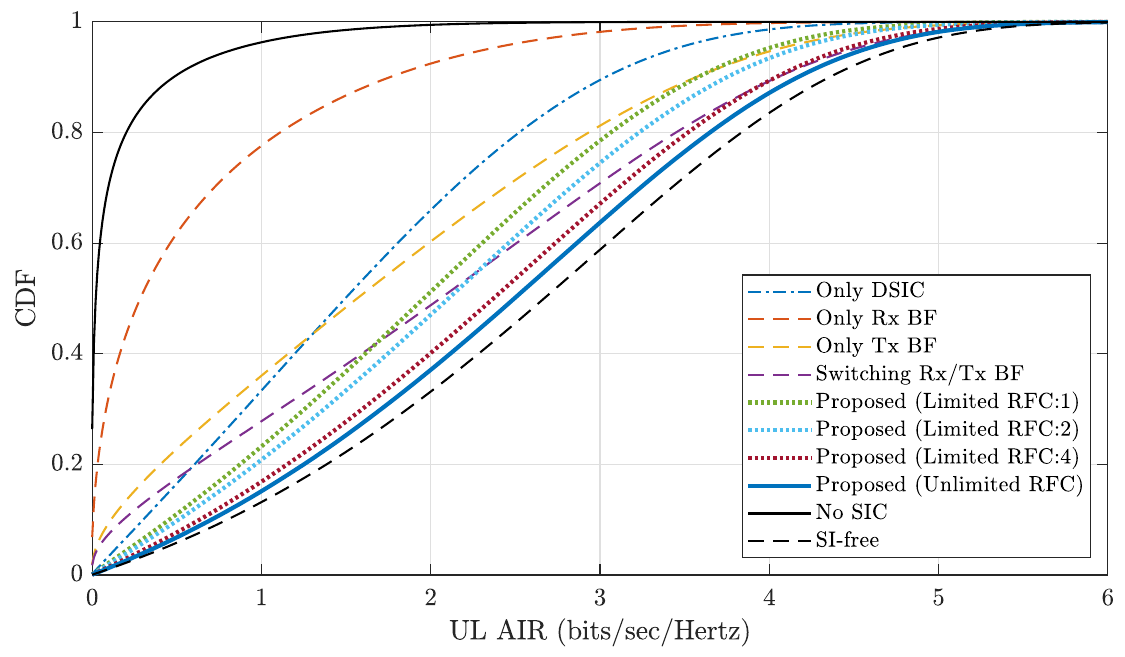}
	}
	\caption{CDF of UL AIR where user and scatterer positions are random. ($K_d=4$, $K_u=4$, $K_s=13$. User and scatterer positions are random except one DL user and one UL user are always at $\theta=0^\circ$ as an FD user.)}
	\label{fig_5}
\end{figure}

\begin{figure}[tb]
\centering
\resizebox{0.45\textwidth}{!}{%
	\includegraphics{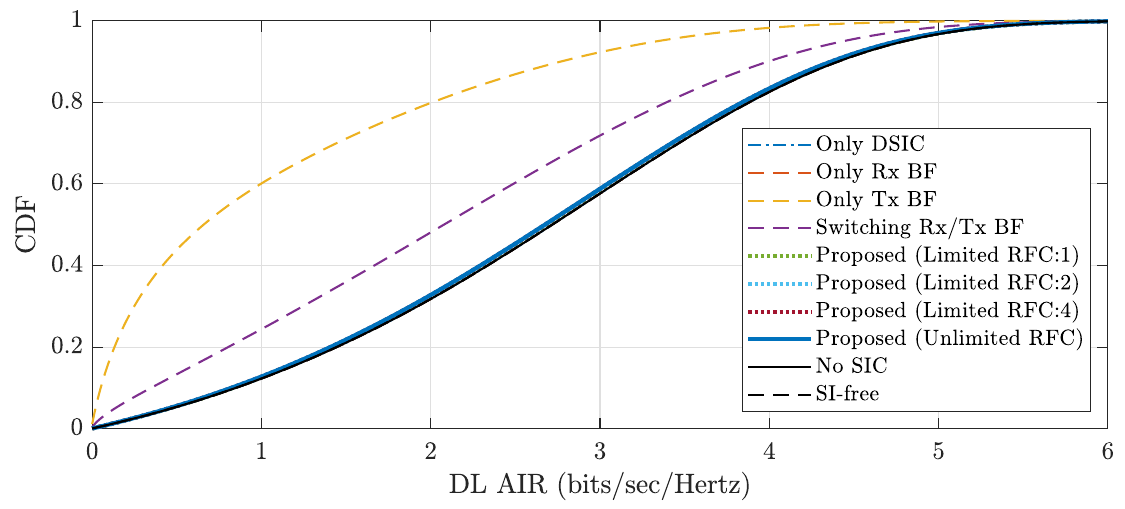}
	}
	\caption{CDF of DL AIR where user and scatterer positions are random. ($K_d=4$, $K_u=4$, $K_s=13$. User and scatterer positions are random except one DL user and one UL user are always at $\theta=0^\circ$ as an FD user.)}
	\label{fig_6}
\end{figure}

For a single case of user settlement and a single scatterer, Figures \ref{fig_3} and \ref{fig_4} showed the basic mechanisms affecting the performance related to user-scatterer positioning. The proposed selection offers a way to choose the most suitable action for a scatterer. In the multi-scatterer case, each scatterer is considered individually and assigned with an action just as in the single-scatterer case. Fig. \ref{fig_9} shows these action assignments in three exemplary scenarios. Five methods are labeled as M1 to M5 in the legend at the right top. Three bottom subplots show the action assignments in three scenarios, according to the legend at the right bottom. Finally, the resultant performances are shown with the bar plot on top. In this plot, the implementation of the proposed selection algorithm in the limited RFC case is given as M3 and M4. This is the case where $L_{Rx}=L_{Tx}<K_s$ and $L_{DSIC}=L_{NA}=K_s$, which are the parameters affecting the Algorithm 1. As seen, even if $L_{Rx}=L_{Tx}<K_s$, the performance is very close to the unlimited case thanks to the metric-based selection algorithm.

The selection mechanism in the multi-scatterer case is illustrated in Fig. \ref{fig_9}, but the performance comparison is still dependent on the given scenarios. To allow more inclusive performance comparison for the multi-scatterer case, the next figures will show results from simulations where user and scatterer positions are random. Figures \ref{fig_5} and \ref{fig_6} show UL and DL performances in terms of the CDF of UL and DL AIR. First of all, DL performance is affected only by the Only Tx BF and Switching Rx/Tx BF cases. On the other hand, it is seen from the CDF of UL AIR that the worst option is to apply only Tx BF or only Rx BF. "Only DSIC" and "Switching Rx/Tx BF" are better but have serious performance losses. The proposed selection strategy performs very closely to the SI-free case even if the RFCs are limited. Note that comparing the limited cases of the proposed selection policy with the Switching Rx/Tx BF case would not be fair because there is no such limitation in the latter case.

\begin{figure}[tb]
\centering
\resizebox{0.45\textwidth}{!}{%
	\includegraphics{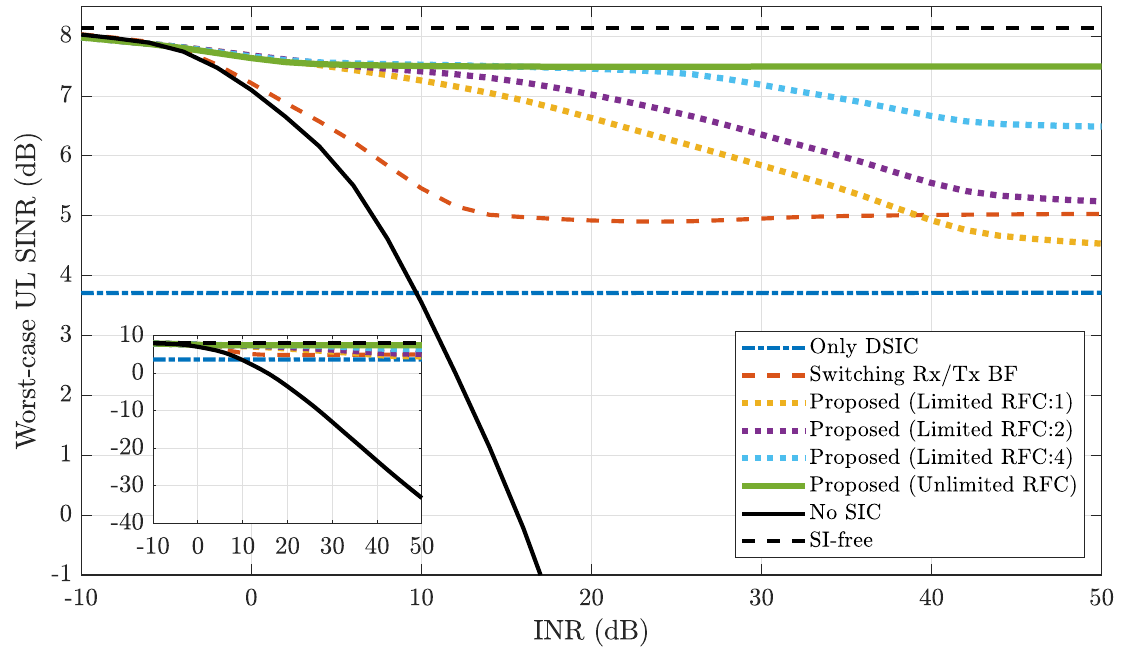}
	}
	\caption{Worst-case UL SINR against INR: $P_k^{(s)}/N_0$. ($K_d=4$, $K_u=4$, $K_s=13$. User and scatterer positions are random except one DL user and one UL user are always at $\theta=0^\circ$ as an FD user.)}
	\label{fig_7}
\end{figure}

\begin{figure}[tb]
\centering
\resizebox{0.45\textwidth}{!}{%
	\includegraphics{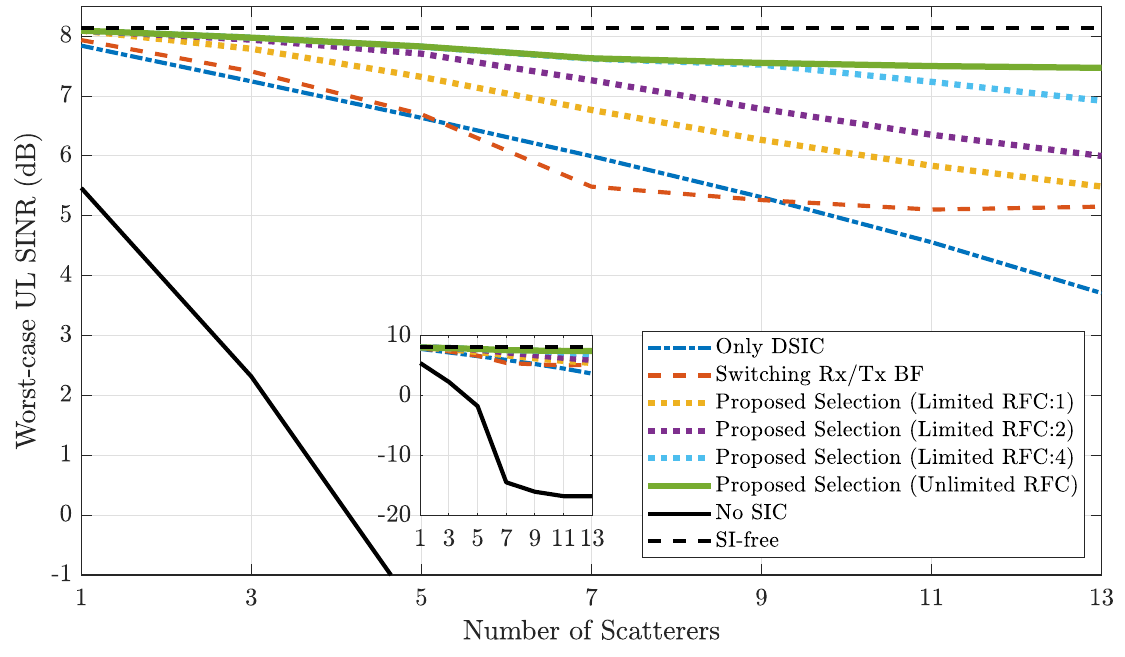}
	}
	\caption{Worst-case UL SINR against the number of scatterers. ($K_d=4$, $K_u=4$. User and scatterer positions are random except one DL user and one UL user are always at $\theta=0^\circ$ as an FD user.)}
	\label{fig_8}
\end{figure}

SI mitigation performance is shown against SI power in Fig. \ref{fig_7}, and against the number of scatterers in Fig. \ref{fig_8}. The most interesting finding is that the DSIC performance depends not on SI power but on the number of scatterers. The reason is that the SI channel is more accurately estimated when the SI is more powerful, and it is erroneous when the SI is weak. However, each scatterer injects its own statistically independent error into the regenerated SI, and the performance monotonically decreases with the number of scatterers. The obvious result is that the DSIC should not be applied for all the scatterers, since it causes worse performance than the No SIC case for weak SI, and each scatterer causes an extra deterioration. The performance decrease in the Switching Rx/Tx BF case against increasing SI power is limited since the BF design accounts for the SI power. This is consistent with the fact that the calculated metrics for BF cases in Table \ref{tab_metrics} are not 0 but $1-Q$ although INR is assumed to go to infinity in the derivation. The relation between the performance in the spatial suppression case and the number of the scatterers is based on the chance that users and scatterers close or not, as discussed in Figures \ref{fig_3} and \ref{fig_4}. On the other hand, the proposed selection policy is superior to these conventional methods, and robust against the number of scatterers and the SI power, even if the number of RFCs is limited to 2 or 4. 

Figures \ref{fig_10} and \ref{fig_11} are about the detection, mitigation, and initial angular estimation in the case of an emerging scatterer, detailed in Section \ref{sec_emerging}. While FD communication continues with the proposed SI mitigation strategy, an unknown scatterer emerges. This scatterer can deteriorate the performance up to 25 dB depending on its angular position as seen in Fig. \ref{fig_10}. Then, the detection mechanism detects the scatterer with the probability of detection shown in the bottom subplot. If detection occurs, DSIC is repeated with the inclusion of the detected scatterer. In the end, the recovery shown in the upper subplot is achieved, which is very close to the case without the new scatterer. Note that the user and scatterer positions are the same as the first scenario shown in Fig. \ref{fig_9}, therefore the actions assigned to each scatterer can be followed from there. It is seen in Fig. \ref{fig_10} that when the new scatterer is close to a known scatterer assigned with beamforming, it does not affect the performance and cannot be detected. Also, if a new scatterer emerges far from the users, it has a limited impact and can be detected rarely. So, the emerging scatterer can be detected if it deteriorates the performance, but an almost perfect recovery is possible afterward. 

After the recovery, the angle of the new scatterer is estimated immediately using the obtained SI channel estimate. The accuracy of the proposed estimator is shown in Fig. \ref{fig_11}. It is seen that the angular error is almost always below 0.5$^\circ$. However, rare cases of gross errors occur outside the beams, preventing blue and yellow curves from reaching one (for example, detections between -30$^\circ$ and -25$^\circ$ in Fig. \ref{fig_10}). Being very accurate except for rare outliers, this initial angular estimate can be beneficial in converting an exhaustive search into a simple verification process.

\section{Conclusion}
This paper investigates the effect of FFSI on FD MIMO communications. It is observed that the FFSI also degrades the performance significantly, while the NFSI is more extensively studied in the literature. For SI mitigation, DSIC and spatial suppression are examined and their disadvantages are highlighted. Our proposed selection policy bypasses both disadvantages by assigning the most suitable action for each scatterer based on derived metrics. Furthermore, models and methods are based on the angle and delay parameters of scatterers, collected in a scatterer map. The maintenance of the scatterer map is addressed, with a particular focus on newly emerging scatterers. It is shown that if an emerging scatterer degrades the performance, besides being detected and mitigated successfully, its delay and angle are accurately estimated to be added to the scatterer map, exhibiting a sensing for communication case.

\appendix[Beamformer Optimizations] \label{append}
 The general subject of the system is to minimize the total SI and noise powers, i.e. $\sum_{k_u=1}^{K_u} \left(\text{var}(z_{k_u,m}^{(si)}) + \text{var}(z_{k_u,m}^{(n)})\right)$ from Section \ref{sec_power_exp}, while maximizing the UL and DL intended signal powers, for $\bm{c}_{k_u}^H \bm{c}_{k_u}=1$ and $\bm{w}_{k_d}^H \bm{w}_{k_d}=1$. To avoid a joint optimization of $\bm{c}_{k_u}$ for $k_u \in \mathcal{U}_u$ and $\bm{w}_{k_d}$ for $k_d \in \mathcal{U}_d$, we can optimize only Rx or Tx side while the other is given. To optimize Tx beamformers for given Rx beamformers, similar to \emph{max-SLNR beamformer} \cite{Sadek07}, the problem is written as
\begin{equation} \label{eq_gen_slnr}
    \max_{\bm{w}_{k_d}} 
    \dfrac{P_{k_d}^{(d)} |\bm{w}_{k_d}^H \bm{q}_{k_d}^{(d)}|^2} 
    {\sum\limits_{k_u=1}^{K_u}\left(\sum\limits_{k_d'=1}^{K_d} \sum\limits_{k_s=1}^{K_s} P_{k_s}^{(s)} |\bm{w}_{k_d'}^H \bm{q}_{k_s}^{(s)}|^2 |\bm{c}_{k_u}^H \bm{q}_{k_s}^{(s)}|^2 +  N_0\right)},
\end{equation}
for $\bm{w}_{k_d}^H \bm{w}_{k_d}=1$, where we implemented $\bm{c}_{k_u}^H \bm{c}_{k_u}=1$. Due to the denominator, joint optimization of all $\bm{w}_{k_d}$ is needed. Instead, we suboptimally assume that the SI contribution from each $\bm{w}_{k_d'}$ for $k_d'\neq k_d$ equals that of $\bm{w}_{k_d}$, which removes the summation over $k_d'$ and multiplies the SI term by $K_d$. Also, we replace $\bm{w}_{k_d}$ by $\bm{w}_{k_d}'/\sqrt{(\bm{w}_{k_d}')^H\bm{w}_{k_d}'}$ to drop $\bm{w}_{k_d}^H \bm{w}_{k_d}=1$ constraint. When necessary cancellations are made and the prime is dropped for notational consistency, we reach
\begin{equation}
    \max_{\bm{w}_{k_d}} 
    \dfrac{|\bm{w}_{k_d}^H \bm{q}_{k_d}^{(d)}|^2} 
    {\sum\limits_{k_s=1}^{K_s} \sum\limits_{k_u=1}^{K_u} P_{k_s}^{(s)} |\bm{w}_{k_d}^H \bm{q}_{k_s}^{(s)}|^2 |\bm{c}_{k_u}^H \bm{q}_{k_s}^{(s)}|^2 + \dfrac{K_u}{K_d} N_0\bm{w}_{k_d}^H \bm{w}_{k_d}},
\end{equation}
whose result is to be scaled to have a unit norm. We can rewrite the problem in a better form as
\begin{equation}
    \max_{\bm{w}_{k_d}} 
    \dfrac{|\bm{w}_{k_d}^H \bm{q}_{k_d}^{(d)}|^2} 
    {\bm{w}_{k_d}^H\left(\sum_{k_s=1}^{K_s} \rho_{k_s}^{(s)} \bm{q}_{k_s}^{(s)} (\bm{q}_{k_s}^{(s)})^{H}+ \dfrac{K_u}{K_d} N_0\bm{I}_{N_a} \right)\bm{w}_{k_d}},
\end{equation}
where $\rho_{k_s}^{(s)}\triangleq P_{k_s}^{(s)} \sum_{k_u=1}^{K_u} |\bm{c}_{k_u}^H \bm{q}_{k_s}^{(s)}|^2$. This particular form of the problem is known to have the solution written in \eqref{eq_opt_w}.
 
 Alternatively, Rx beamformers can be optimized for given Tx beamformers. In this case, we focus separately on their individual outputs as
\begin{equation}
     \max_{\bm{c}_{k_u}} 
    \dfrac{P_{k_u}^{(u)} |\bm{c}_{k_u}^H \bm{q}_{k_u}^{(u)}|^2}
    {\sum\limits_{k_s=1}^{K_s} \sum\limits_{k_d=1}^{K_d} P_{k_s}^{(s)} |\bm{w}_{k_d}^H \bm{q}_{k_s}^{(s)}|^2 |\bm{c}_{k_u}^H \bm{q}_{k_s}^{(s)}|^2 + N_0 \bm{c}_{k_u}^H \bm{c}_{k_u}},
\end{equation}
to be solved for each $k_u$ separately. The metric to be maximized is SINR and the optimal solution is called the \emph{max-SINR beamformer}. We can rewrite the problem as
\begin{equation}
    \max_{\bm{c}_{k_u}} 
    \dfrac{|\bm{c}_{k_u}^H \bm{q}_{k_u}^{(u)}|^2}
    {\bm{c}_{k_u}^H \left(\sum_{k_s=1}^{K_s} \rho_{k_s}^{(s)} \bm{q}_{k_s}^{(s)} (\bm{q}_{k_s}^{(s)})^{H} + N_0 \bm{I}_{N_a}\right) \bm{c}_{k_u}},
\end{equation}
where $\rho_{k_s}^{(s)} \triangleq P_{k_s}^{(s)} \sum_{k_d=1}^{K_d} |\bm{w}_{k_d}^H \bm{q}_{k_s}^{(s)}|^2$. This is a well-known optimization problem with the solution written in \eqref{eq_opt_c}.

\bibliography{references} 

\begin{thebibliography}{10}
\providecommand{\url}[1]{#1}
\csname url@samestyle\endcsname
\providecommand{\newblock}{\relax}
\providecommand{\bibinfo}[2]{#2}
\providecommand{\BIBentrySTDinterwordspacing}{\spaceskip=0pt\relax}
\providecommand{\BIBentryALTinterwordstretchfactor}{4}
\providecommand{\BIBentryALTinterwordspacing}{\spaceskip=\fontdimen2\font plus
\BIBentryALTinterwordstretchfactor\fontdimen3\font minus
  \fontdimen4\font\relax}
\providecommand{\BIBforeignlanguage}[2]{{%
\expandafter\ifx\csname l@#1\endcsname\relax
\typeout{** WARNING: IEEEtran.bst: No hyphenation pattern has been}%
\typeout{** loaded for the language `#1'. Using the pattern for}%
\typeout{** the default language instead.}%
\else
\language=\csname l@#1\endcsname
\fi
#2}}
\providecommand{\BIBdecl}{\relax}
\BIBdecl

\bibitem{Kurt_pimrc}
A.~Kurt and G.~M. Guvensen, ``Sensing and mitigation of far-field
  self-interference for full-duplex {MIMO} systems,'' in \emph{IEEE 35th Annu.
  Int. Symp. Pers. Indoor Mobile Radio Commun. (PIMRC)}, 2024.

\bibitem{Mohammadi_TenYears}
M.~Mohammadi, Z.~Mobini, H.~Q. Ngo, and M.~Matthaiou, ``Ten years of research
  advances in full-duplex massive {MIMO},'' \emph{IEEE Trans. Commun.}, pp.
  1--1, 2024.

\bibitem{Liu_FD_ISAC}
Z.~Liu, S.~Aditya, H.~Li, and B.~Clerckx, ``Joint transmit and receive
  beamforming design in full-duplex integrated sensing and communications,''
  \emph{IEEE J. Sel. Areas Commun.}, vol.~41, no.~9, pp. 2907--2919, 2023.

\bibitem{smida23}
B.~Smida, A.~Sabharwal, G.~Fodor, G.~C. Alexandropoulos, H.~A. Suraweera, and
  C.-B. Chae, ``Full-duplex wireless for {6G}: Progress brings new
  opportunities and challenges,'' \emph{IEEE J. Sel. Areas Commun.}, vol.~41,
  no.~9, pp. 2729--2750, 2023.

\bibitem{Sheemar_Multicell}
C.~K. Sheemar, S.~Chatzinotas, D.~Slock, E.~Lagunas, and J.~Querol, ``Parallel
  and distributed hybrid beamforming for multicell millimeter wave {MIMO} full
  duplex,'' \emph{IEEE Trans. Veh. Technol.}, pp. 1--16, 2024.

\bibitem{Sabharwal_Challenges}
A.~Sabharwal, P.~Schniter, D.~Guo, D.~W. Bliss, S.~Rangarajan, and R.~Wichman,
  ``In-band full-duplex wireless: Challenges and opportunities,'' \emph{IEEE J.
  Sel. Areas Commun.}, vol.~32, no.~9, pp. 1637--1652, 2014.

\bibitem{Kim_PHY_MAC}
D.~Kim, H.~Lee, and D.~Hong, ``A survey of in-band full-duplex transmission:
  From the perspective of {PHY} and {MAC} layers,'' \emph{IEEE Commun. Surv.
  Tut.}, vol.~17, no.~4, pp. 2017--2046, 2015.

\bibitem{Sattari_Estimation}
M.~Sattari, H.~Guo, D.~Gündüz, A.~Panahi, and T.~Svensson, ``Full-duplex
  millimeter wave {MIMO} channel estimation: A neural network approach,''
  \emph{IEEE Trans. Mach. Learn. Commun. Netw.}, vol.~2, pp. 1093--1108, 2024.

\bibitem{Roberts_Codebooks}
I.~P. Roberts, H.~B. Jain, S.~Vishwanath, and J.~G. Andrews, ``Millimeter wave
  analog beamforming codebooks robust to self-interference,'' in \emph{2021
  IEEE Global Commun. Conf. (GLOBECOM)}, 2021, pp. 1--6.

\bibitem{Roberts22}
I.~P. Roberts, A.~Chopra, T.~Novlan, S.~Vishwanath, and J.~G. Andrews,
  ``Beamformed self-interference measurements at 28 {GHz}: Spatial insights and
  angular spread,'' \emph{IEEE Trans. Wireless Commun.}, vol.~21, no.~11, pp.
  9744--9760, 2022.

\bibitem{Barneto_FDSensing}
C.~B. Barneto, T.~Riihonen, S.~D. Liyanaarachchi, M.~Heino,
  N.~González-Prelcic, and M.~Valkama, ``Beamformer design and optimization
  for joint communication and full-duplex sensing at {mm-Waves},'' \emph{IEEE
  Trans. Commun.}, vol.~70, no.~12, pp. 8298--8312, 2022.

\bibitem{Everett14}
E.~Everett, A.~Sahai, and A.~Sabharwal, ``Passive self-interference suppression
  for full-duplex infrastructure nodes,'' \emph{IEEE Trans. Wireless Commun.},
  vol.~13, no.~2, pp. 680--694, 2014.

\bibitem{Chen_propagation}
Y.~Chen, C.~Ding, Y.~Jia, and Y.~Liu, ``Antenna/propagation domain
  self-interference cancellation ({SIC}) for in-band full-duplex wireless
  communication systems,'' \emph{Sensors}, vol.~22, no.~5, 2022.

\bibitem{Ding_Movable}
J.~Ding, Z.~Zhou, W.~Li, C.~Wang, L.~Lin, and B.~Jiao, ``Movable
  antenna-enabled co-frequency co-time full-duplex wireless communication,''
  \emph{IEEE Commun. Lett.}, vol.~28, no.~10, pp. 2412--2416, 2024.

\bibitem{Askar_Handling}
R.~Askar, J.~Chung, Z.~Guo, H.~Ko, W.~Keusgen, and T.~Haustein, ``Interference
  handling challenges toward full duplex evolution in {5G} and beyond cellular
  networks,'' \emph{IEEE Wireless Commun.}, vol.~28, no.~1, pp. 51--59, 2021.

\bibitem{Le_BeamBasedASIC}
A.~T. Le, L.~C. Tran, X.~Huang, and Y.~J. Guo, ``Beam-based analog
  self-interference cancellation in full-duplex {MIMO} systems,'' \emph{IEEE
  Trans. Wireless Commun.}, vol.~19, no.~4, pp. 2460--2471, 2020.

\bibitem{Kwak_ASIC}
J.~W. Kwak, M.~S. Sim, I.-W. Kang, J.~Park, K.-K. Wong, and C.-B. Chae,
  ``Analog self-interference cancellation with practical {RF} components for
  full-duplex radios,'' \emph{IEEE Trans. Wireless Commun.}, vol.~22, no.~7,
  pp. 4552--4564, 2023.

\bibitem{Kim_Duplexer}
H.~Kim, K.~Ko, K.~Kwon, I.-C. Hwang, and S.~Park, ``Fast electrical balance
  duplexer tuning using neural networks for {RF} self-interference cancellation
  in in-band full-duplex systems,'' \emph{IEEE Access}, vol.~12, pp.
  151\,805--151\,824, 2024.

\bibitem{Soriano_Adaptive}
F.~J. Soriano-Irigaray, J.~S. Fernandez-Prat, F.~J. Lopez-Martinez,
  E.~Martos-Naya, O.~Cobos-Morales, and J.~T. Entrambasaguas, ``Adaptive
  self-interference cancellation for full duplex radio: Analytical model and
  experimental validation,'' \emph{IEEE Access}, vol.~6, pp. 65\,018--65\,026,
  2018.

\bibitem{Kiayani_NLRF}
A.~Kiayani, M.~Z. Waheed, L.~Anttila, M.~Abdelaziz, D.~Korpi, V.~Syrjälä,
  M.~Kosunen, K.~Stadius, J.~Ryynänen, and M.~Valkama, ``Adaptive nonlinear
  {RF} cancellation for improved isolation in simultaneous transmit–receive
  systems,'' \emph{IEEE Trans. Microw. Theory Techn.}, vol.~66, no.~5, pp.
  2299--2312, 2018.

\bibitem{Liu_ASICwImperfect}
D.~Liu, Y.~Shen, S.~Shao, Y.~Tang, and Y.~Gong, ``On the analog
  self-interference cancellation for full-duplex communications with imperfect
  channel state information,'' \emph{IEEE Access}, vol.~5, pp. 9277--9290,
  2017.

\bibitem{Liu_dig_assisted}
Y.~Liu, X.~Quan, W.~Pan, and Y.~Tang, ``Digitally assisted analog interference
  cancellation for in-band full-duplex radios,'' \emph{IEEE Commun. Lett.},
  vol.~21, no.~5, pp. 1079--1082, 2017.

\bibitem{kurt23}
A.~Kurt, M.~B. Salman, U.~B. Sarac, and G.~M. Guvensen, ``An adaptive-iterative
  nonlinear interference cancellation in time-varying full-duplex channels,''
  \emph{IEEE Trans. Veh. Technol.}, vol.~72, no.~2, pp. 1862--1878, 2023.

\bibitem{Campo_Spline}
P.~Pascual~Campo, L.~Anttila, D.~Korpi, and M.~Valkama, ``Cascaded spline-based
  models for complex nonlinear systems: Methods and applications,'' \emph{IEEE
  Trans. Signal Process.}, vol.~69, pp. 370--384, 2021.

\bibitem{Elsayed_ML}
M.~Elsayed, A.~A.~A. El-Banna, O.~A. Dobre, W.~Y. Shiu, and P.~Wang, ``Machine
  learning-based self-interference cancellation for full-duplex radio:
  Approaches, open challenges, and future research directions,'' \emph{IEEE
  Open J. Veh. Technol.}, vol.~5, pp. 21--47, 2024.

\bibitem{Everett_SoftNull}
E.~Everett, C.~Shepard, L.~Zhong, and A.~Sabharwal, ``Softnull: Many-antenna
  full-duplex wireless via digital beamforming,'' \emph{IEEE Trans. Wireless
  Commun.}, vol.~15, no.~12, pp. 8077--8092, 2016.

\bibitem{kim23}
S.-M. Kim, Y.-G. Lim, L.~Dai, and C.-B. Chae, ``Performance analysis of
  self-interference cancellation in full-duplex massive {MIMO} systems:
  Subtraction versus spatial suppression,'' \emph{IEEE Trans. Wireless
  Commun.}, vol.~22, no.~1, pp. 642--657, 2023.

\bibitem{Balti_HBFDesign}
E.~Balti, S.~Akoum, I.~Alfalujah, and B.~L. Evans, ``Hybrid beamforming design
  for full-duplex millimeter wave massive {MIMO} systems,'' \emph{IEEE Trans.
  Veh. Technol.}, vol.~73, no.~11, pp. 17\,041--17\,058, 2024.

\bibitem{koc21}
A.~Koc and T.~Le-Ngoc, ``Full-duplex {mmWave} massive {MIMO} systems: A joint
  hybrid precoding/combining and self-interference cancellation design,''
  \emph{IEEE Open J. Commun. Soc.}, vol.~2, pp. 754--774, 2021.

\bibitem{Keskin_Monostatic_PN}
M.~F. Keskin, H.~Wymeersch, and V.~Koivunen, ``Monostatic sensing with {OFDM}
  under phase noise: From mitigation to exploitation,'' \emph{IEEE Trans.
  Signal Process.}, vol.~71, pp. 1363--1378, 2023.

\bibitem{Liu_ISAC_toward}
F.~Liu, Y.~Cui, C.~Masouros, J.~Xu, T.~X. Han, Y.~C. Eldar, and S.~Buzzi,
  ``Integrated sensing and communications: Toward dual-functional wireless
  networks for {6G} and beyond,'' \emph{IEEE J. Sel. Areas Commun.}, vol.~40,
  no.~6, pp. 1728--1767, 2022.

\bibitem{Chen_Sparsity}
Z.~Chen and C.~Yang, ``Pilot decontamination in wideband massive {MIMO} systems
  by exploiting channel sparsity,'' \emph{IEEE Trans. Wireless Commun.},
  vol.~15, no.~7, pp. 5087--5100, 2016.

\bibitem{Rangan_Sparsity}
S.~Rangan, T.~S. Rappaport, and E.~Erkip, ``Millimeter-wave cellular wireless
  networks: Potentials and challenges,'' \emph{Proc. IEEE}, vol. 102, no.~3,
  pp. 366--385, 2014.

\bibitem{Sloane_Sparsity}
W.~Sloane, C.~Gentile, M.~Shafi, J.~Senic, P.~A. Martin, and G.~K. Woodward,
  ``Measurement-based analysis of millimeter-wave channel sparsity,''
  \emph{IEEE Antennas Wireless Propag. Lett.}, vol.~22, no.~4, pp. 784--788,
  2023.

\bibitem{Kay93}
S.~M. Kay, \emph{Fundamentals of Statistical Signal Processing: Estimation
  Theory}.\hskip 1em plus 0.5em minus 0.4em\relax USA: Prentice-Hall, Inc.,
  1993.

\bibitem{Sadek07}
M.~Sadek, A.~Tarighat, and A.~H. Sayed, ``A leakage-based precoding scheme for
  downlink multi-user {MIMO} channels,'' \emph{IEEE Trans. Wireless Commun.},
  vol.~6, no.~5, pp. 1711--1721, 2007.

\bibitem{book_richards}
M.~Richards, \emph{Fundamentals of Radar Signal Processing, Second
  Edition}.\hskip 1em plus 0.5em minus 0.4em\relax McGraw Hill LLC, 2013.

\end{thebibliography}
\bibliographystyle{IEEEtran}

\vfill

\end{document}